\documentclass[aps,showpacs,nofootinbib,eqsecnum]{revtex4-2}

\usepackage{amsmath,amssymb}
\usepackage{bm}
\usepackage{cancel}
\usepackage{cases}
\usepackage{color}
\usepackage{comment}
\usepackage{dcolumn}
\usepackage{graphicx}
\usepackage{here}
\usepackage{multirow}
\usepackage{slashed}
\usepackage{subfigure}
\usepackage{textcase}
\usepackage{tikz-feynhand}
\usepackage{ytableau}
\usepackage[mathlines]{lineno}
\usepackage[dvipsnames]{xcolor}
\usepackage[colorlinks=true,pdfstartview=FitV,linkcolor=blue,citecolor=blue,urlcolor=blue]{hyperref}

\begin{document}

\begin{flushright}
    KOBE-COSMO-26-12, KUNS-3131
\end{flushright} 

\title{
Photon evolution through light-particle conversion in stochastic magnetic fields:
\\
Kramers--Kronig relations and universality
}

\author{Wataru Chiba$^{1}$}
\thanks{{\color{blue}245s114s@stu.kobe-u.ac.jp}}

\author{Ryusuke Jinno$^{1}$}
\thanks{{\color{blue}jinno@phys.sci.kobe-u.ac.jp}}

\author{Kimihiro Nomura$^{2}$}
\thanks{{\color{blue}k.nomura@tap.scphys.kyoto-u.ac.jp}}

\affiliation{$^{1}$ Department of Physics, Graduate School of Science, Kobe University, 1-1 Rokkodai, Kobe, Hyogo 657-8501, Japan}

\affiliation{$^{2}$ Department of Physics, Kyoto University, Kyoto 606-8502, Japan}

\date{\today}

\begin{abstract}
We study photon conversion into light particles, such as dilatons, axions, and gravitons, induced by background magnetic fields, and formulate the resulting evolution in terms of the Stokes parameters.
We show that photon conversion reduces to an effective non-unitary evolution of the photon sector.
In the perturbative regime, the Hermitian and anti-Hermitian parts of the effective photon Hamiltonian are related through Kramers--Kronig relations, revealing that the changes in the Stokes parameters are not independent.
We apply this general framework to photon evolution in Gaussian stochastic magnetic fields with possible helicity, and derive the ensemble-averaged evolution of the Stokes parameters.
We find that two nonzero contributions arise even in the absence of magnetic helicity, which respectively correspond to the overall attenuation of all the Stokes parameters and to an irrelevant overall phase.
We also find that magnetic helicity gives rise to two additional contributions, one of which changes the circular polarization while the other induces a rotation of linear polarization.
These contributions form two Kramers--Kronig pairs.
In particular, photon conversion induced by magnetic helicity is accompanied by both circular-polarization and birefringent responses.
The same structure appears in dilaton-photon, axion-photon, and graviton-photon systems, revealing universality of photon conversion in magnetic fields.

\end{abstract}

\maketitle

\section{Introduction}

The conversion of photons into other light particles in an external magnetic field provides a powerful probe of physics beyond the Standard Model.
Representative examples include axion-photon conversion through the Primakoff effect~\cite{MAIANI1986359,PhysRevD.37.1237} and graviton-photon conversion through the Gertsenshtein effect~\cite{gertsenshtein1962wave,PhysRevD.37.1237}.
These conversion phenomena have been extensively studied as experimental and observational probes of axions and axion-like particles~\cite{Jain_2002,Mirizzi_2017,PhysRevLett.51.1415,Irastorza:2018dyq,CAST:2017uph,Armengaud:2014gea,Asztalos_2010,ParticleDataGroup:2024cfk}, dilatons and other light scalar fields~\cite{Chaubey_2023,chaubey2022differentiatingdilatonsaxionsmixing}, as well as high-frequency gravitational waves originating from the early Universe and primordial black hole binaries~\cite{he2024inversegertsenshteineffectprobe,Ito:2023fcr,Kushwaha:2025mia,Gupta_2026}.
Describing the evolution of the photon state in a magnetic field is therefore essential for extracting information about such weakly interacting degrees of freedom.

Photon conversion into additional degrees of freedom, such as dilatons, axions, and gravitons, possesses a common underlying structure, in which the conversion is induced by a background magnetic field.
In these systems, the photon and the additional degrees of freedom evolve together as a closed system, whereas the photon sector alone undergoes an effective open-system evolution as photons convert into the other particles.
This perspective allows us to formulate photon evolution in a unified description as an open system, independently of the specific nature of the conversion channel.

In this work, we develop such a description by projecting the full dynamics onto the photon sector and characterize the general structure of the effective photon evolution induced by conversion.
We formulate this evolution directly in terms of the Stokes parameters, which characterize both the intensity and polarization state of photons.
When photons are weakly coupled to additional fields, the evolution of the photon sector is governed by an effective non-Hermitian Hamiltonian, which in the perturbative regime leads to a linear transformation of the Stokes parameters.
We derive the general structure of this transformation and show that its transformation parameters are determined by the Hermitian and anti-Hermitian parts of the effective photon Hamiltonian.
In the perturbative regime, these two parts are not independent but are related through Kramers--Kronig (KK) relations as a consequence of the causal structure of the photon evolution.
We then explicitly determine the transformation parameters for three representative systems: photon-dilaton, photon-axion, and photon-graviton mixing in an external magnetic field.
This formulation extends our previous studies~\cite{Chiba:2025hgr,Chiba:2025odu} of the conversion of initially unpolarized photons into axions or gravitons to arbitrary initial polarization states.
More importantly, it reveals a common structure underlying the effective photon evolution across these conversion channels.

We further apply this framework to photon evolution in a stochastic magnetic-field background with Gaussian statistics characterized by its power spectra including both nonhelical and helical components.
This setup provides a natural model for observational settings in which photons propagate through cosmological magnetic fields while undergoing partial conversion into other degrees of freedom.
We consider the ensemble-averaged evolution of the photon observables and show that the expectation values of the Stokes parameters can be expressed as convolutions of the magnetic-field power spectra with kernels that encode the photon-particle mixing.
Remarkably, two nontrivial results emerge after ensemble averaging over the stochastic magnetic field.
First, we find that the resulting convolution kernels take the same functional form for the photon-dilaton, photon-axion, and photon-graviton systems.
This universality provides a unified description of the averaged photon evolution, despite the different conversion channels.
Second, in the presence of magnetic helicity, two characteristic contributions survive the ensemble average: one changes the circular polarization, while the other produces birefringence.
Crucially, these two contributions are not independent, but constitute a KK pair.
Consequently, the change in circular polarization and birefringence induced by magnetic helicity necessarily appear together, revealing a direct connection between two effects that might otherwise appear unrelated.

The present paper is organized as follows.
In Sec.~\ref{sec:Stokes}, we formulate the evolution of the photon Stokes parameters under a general non-unitary evolution.
In Sec.~\ref{sec:conversion}, we show how such an effective non-unitary evolution arises from photon conversion into additional degrees of freedom and, in the perturbative regime, derive the transformation parameters and their KK relations.
In Sec.~\ref{sec:magnetic}, we apply the general framework to specific cases of photon-particle mixing in an external magnetic field.
In particular, we explicitly derive the transformation parameters for photon mixing with scalar (dilaton), pseudoscalar (axion), and tensor (graviton) degrees of freedom.
In Sec.~\ref{sec:random}, we introduce stochastic magnetic fields and study the evolution of the Stokes parameters in these backgrounds.
Assuming that the magnetic field is Gaussian and characterized by its power spectra, we represent the evolution of the full system to all orders in the interaction and the evolution of the ensemble-averaged Stokes parameters.
In Sec.~\ref{sec:numerical}, adopting a specific magnetic-field spectrum, we numerically study the changes in the Stokes parameters induced by photon-particle conversion.
Finally, Sec.~\ref{sec:discussions} is devoted to discussion and conclusions.

\section{General evolution of photon Stokes parameters}
\label{sec:Stokes}

In this section, we first review the description of photon intensity and polarization in terms of Stokes parameters.
We then describe the evolution of the photon intensity and polarization with a general non-Hermitian matrix as a linear transformation among the Stokes parameters.

\subsection{Photon propagation}

Throughout this paper, we take photons to propagate along the $z$ direction and focus on the two transverse components $A_x$ and $A_y$ in the $x$-$y$ plane.
To derive the propagation equation, we start with the Maxwell Lagrangian in vacuum,
\begin{align}
    \mathcal{L}_{0}=-\frac{1}{4}F_{\mu\nu}F^{\mu\nu}~,
\end{align}
where $F_{\mu\nu}=\partial_\mu A_\nu-\partial_\nu A_\mu$ and $A_\mu$ is the photon field.
We impose the radiation gauge condition,
\begin{align}
    \partial_i A^i=0~,\quad A^0=0~,\label{eq:gauge}
\end{align}
and introduce the orthonormal basis vectors $e^{(m)}_i$ $(m=x,y,z)$ as 
\begin{align}
    e_i^{(x)}=
    \begin{pmatrix}
        1\\
        0\\
        0\\
    \end{pmatrix}
    ~,\quad
     e_i^{(y)}=
    \begin{pmatrix}
        0\\
        1\\
        0\\
    \end{pmatrix}
    ~,\quad
    e_i^{(z)}=
    \begin{pmatrix}
        0\\
        0\\
        1
    \end{pmatrix}
    ~.
\end{align}
For photons propagating along the $z$ direction, we have $A_z = 0$, and the photon field can be expanded as 
\begin{align}
    A_i(t, \bm{x})=\sum_{m=x,y}A_m(t,z) \, e_i^{(m)}~.
\end{align}
With this notation, the Maxwell Lagrangian reduces to
\begin{align}
    \mathcal{L}_0=\frac{1}{2}\sum_{m=x,y}(\dot{A}_m^2-A_m^{\prime 2})~,
\end{align}
where overdots and primes denote derivatives with respect to time and $z$, respectively, i.e., $\dot{A}\equiv\partial_0 A$ and $A^\prime\equiv\partial_z A$.

We decompose the photon field into monochromatic modes with angular frequency $\omega$ as
\begin{align}
     A_m (t,z)= A_m(z) \, e^{-i\omega t} + \text{c.c.}~,
\end{align}
and derive the equation of motion for the photon field from the above reduced Lagrangian as
\begin{align}
    (-\partial_0^2+\partial_z^2)A_m\equiv\partial^2 A_m=0~.
\end{align}
Adopting a relativistic approximation, the second-order derivative can be approximated as $\partial^2 = (\omega - i\partial_z) (\omega + i\partial_z) \approx 2 \omega(\omega + i\partial_z)$.
Under this approximation, the equation of motion reduces to a first-order differential equation
\begin{align}
    i\partial_z A_m=-\omega A_m\equiv\Pi_{\gamma_m}(\omega) \, A_m~,
\end{align}
where $\Pi_{\gamma_m}$ encodes the dispersion relation of the photon mode $A_m$.
We refer to the first-order equation obtained under this approximation as the propagation equation.
Throughout this work, we consider the propagation of photons in vacuum.
Accordingly, the two photon polarization modes have the same dispersion relation, $\Pi_\gamma\equiv\Pi_{\gamma_x}=\Pi_{\gamma_y}=-\omega$, and do not mix with each other.
 
For a particle with a sufficiently small mass $m$ such that the relativistic approximation remains valid, the differential operator can be approximated as
\begin{align}
(\partial^2-m^2)\simeq2\omega \left( \omega+i\partial_z-\frac{m^2}{2\omega} \right)~.
\end{align}
Accordingly, the corresponding term in the first-order propagation equation is given by
\begin{align}
    \Pi(\omega)=-\omega+\frac{m^2}{2\omega}~.
\end{align}
We later apply this relativistic approximation to light fields such as dilatons and axions.

\subsection{Evolution of the photon Stokes parameters}
\label{subsec:EvolutionStokesParameters}

The intensity and polarization state of the photons are characterized by bilinear combinations of the complex amplitudes $\mathbf{A}=(A_x,A_y)^T$, or equivalently by the $2\times2$ Hermitian photon density matrix $\rho_\gamma\equiv\mathbf{A}\otimes \mathbf{A}^\dagger$, given by
\begin{align}
     \rho_\gamma=
    \begin{pmatrix}
  A_xA_x^*&A_xA_y^*\\A_y A_x^*&A_yA_y^*
\end{pmatrix}
    \equiv \frac12
    \begin{pmatrix}
  I+Q&U-iV\\U+iV&I-Q
\end{pmatrix}
=\frac12 S_\mu\sigma_\mu
~,
\end{align}
where $\sigma_\mu$ collectively denote the identity and Pauli matrices,
\begin{align}
    \sigma_0=
    \begin{pmatrix}
    1&0\\0&1    
    \end{pmatrix}
    ,\qquad
    \sigma_1=
    \begin{pmatrix}
    0&1\\1&0    
    \end{pmatrix}
    ,\qquad
    \sigma_2=
    \begin{pmatrix}
    0&-i\\i&0    
    \end{pmatrix}
    ,\qquad
    \sigma_3=
    \begin{pmatrix}
    1&0\\0&-1    
    \end{pmatrix}
    ~.
\end{align}
The real coefficients $S_\mu=(I,U,V,Q)$ are the Stokes parameters, where $I$ is the total photon intensity, $Q$ and $U$ characterize linear polarization, and $V$ characterizes circular polarization.

For a single field realization, the determinant of $\rho_\gamma$ vanishes identically by definition.
In terms of the Stokes parameters, this implies
\begin{align}
    I^2=Q^2+U^2+V^2~.
\end{align}
Thus, each field realization corresponds to a fully polarized photon state.
More generally, the photon amplitudes may be regarded as classical random variables, or equivalently the density matrix may be understood as an ensemble-averaged quantity.
In that case, the above relation is replaced by the inequality
\begin{align}
    I^2\geq Q^2+U^2+V^2\geq 0~.
\end{align}
This inequality allows photon states to be partially polarized.
In particular, when $Q=U=V=0$, the state is completely unpolarized.

Let us consider a transformation of the photon amplitudes written as $\mathbf{A}_0 \to \mathbf{A}= M\mathbf{A}_0,$ where $\mathbf{A}_0$ is the initial photon amplitude vector and $M$ is a $2\times2$ complex matrix.
The corresponding density matrix transforms as
\begin{align}
\rho_{\gamma 0 } ~ \to ~
\rho_\gamma= M\rho_{\gamma0} M^\dagger~,
\end{align}
where $\rho_{\gamma0}$ is the initial photon density matrix.
The most general linear transformation is given by $M\in GL(2,\mathbb{C})$, which is generated by the algebra $\mathfrak{gl}(2,\mathbb{C})$.
We parameterize the transformation as
\begin{align}
M=e^{-iX},\quad   \mathfrak{gl}(2,\mathbb{C})\ni X=\frac12(j_\mu-ik_\mu)\sigma_\mu~,
\label{eq:algebra}
\end{align}
where the non-Hermitian matrix $X$ generates the linear evolution of the photon amplitude.
This general transformation is characterized by eight real parameters $j_\mu$ and $k_\mu$.

When the parameters $j_\mu$ and $k_\mu$ are sufficiently small, the evolution of the photon density matrix can be expressed as an infinitesimal transformation,
\begin{align}
    \delta \rho_\gamma\equiv\rho_\gamma-\rho_{\gamma0}=-\frac{k_\mu}{2}\{\sigma_\mu,\rho_{\gamma0}\}-i\,\frac {j_\mu}{2} [\sigma_\mu,\rho_{\gamma0}] ~,
    \label{eq:matrix}
\end{align}
where $\{A,B\}\equiv AB+BA$ and $[A,B] \equiv AB - BA$.
When $k_\mu=0$, the first term vanishes, in which case the generator $X$ becomes Hermitian.
Consequently, the evolution is unitary, and the total intensity is conserved.
Thus, the first term represents the non-unitary contribution associated with the open-system nature of the photon sector, whereas the second term describes the unitary evolution generated by the Hermitian part.
Note that $j_0$ does not appear in the evolution of the Stokes parameters because $[\sigma_0,\rho_{\gamma0}]=0$.
Furthermore, if the initial photon state is unpolarized, $\rho_{\gamma0}=(I_0/2)\sigma_0$, it commutes with all the Pauli matrices, $[\sigma_i,\rho_{\gamma0}]=0$.
Consequently, the parameters $j_i$ do not contribute to the evolution of the Stokes parameters for an initially unpolarized state.

This algebra can be decomposed as $\mathfrak{gl}(2,\mathbb{C})\simeq \mathbb{C}I_{2\times2}\oplus\mathfrak{sl}(2,\mathbb{C})$, where $j_0$ and $k_0$ are associated with the real and imaginary parts of the $\mathbb{C}I_{2\times2}$ component of $X$, respectively.
The real component $j_0$ generates an overall phase through the factor $-iX$ in $M=e^{-iX}$, and therefore cancels out in the density matrix.
In contrast, the imaginary component $k_0$ contributes to the real part of $-iX$, changing the overall magnitude of the photon amplitude and thereby uniformly rescaling the Stokes parameters.
Therefore, after removing the physically irrelevant overall phase generated by $j_0$, the algebra acting nontrivially on the Stokes parameters reduces to $\mathbb{R}I_{2\times2}\oplus\mathfrak{sl}(2,\mathbb{C}),$ which is parameterized by seven real parameters: $k_0, k_i,$ and $j_i$ $(i=1,2,3)$.

\section{Photon conversion and effective evolution of Stokes parameters}
\label{sec:conversion}

In this section, we show how an effective non-Hermitian evolution arises when photons can convert into additional degrees of freedom.
By focusing only on the photon sector and treating the additional fields as unobserved degrees of freedom, we can regard the photon sector as an open system, whose evolution is effectively described by a non-Hermitian Hamiltonian.
In Sec.~\ref{subsec:conversion}, we derive this effective photon evolution by eliminating the additional fields from the full system.
In Sec.~\ref{subsec:born}, focusing on the perturbative regime within the Born approximation, we express the eight transformation parameters introduced in Sec.~\ref{sec:Stokes} in terms of the second-order contributions of the full evolution operator.
Finally, in Sec.~\ref{subsec:KK}, we show that the causal structure of these second-order contributions implies Kramers--Kronig relations, which pair the Hermitian and anti-Hermitian parts of the effective photon evolution.

\subsection{Stokes-parameter evolution in conversion dynamics}
\label{subsec:conversion}

Now, we consider a system in which the photon is coupled to additional degrees of freedom.
By focusing on the photon sector of the total system, its effective evolution can be described by the general non-Hermitian framework introduced above.
Suppose that the two photon modes are coupled to an $N$-component external sector $\varphi\equiv(\varphi_1,\cdots,\varphi_N)^T$.
The total system is then described by the amplitude vector $\Psi_{\varphi\gamma}\equiv(\varphi_1,\cdots,\varphi_N,A_x,A_y)^T$, which consists of the external sector and the photon sector.
The density matrix of the total system is defined by $\rho \equiv \Psi_{\varphi\gamma} \otimes \Psi_{\varphi\gamma}^\dagger$.
We assume that the total system evolves unitarily, while the photon and external-sector amplitudes mix during propagation.

In this study, we are interested in the situation in which the initial state at $z=0$ contains only photons, while the external sector has vanishing initial amplitudes.
Therefore, the initial density matrix can be written in block form as 
\begin{align}
    \rho_0\equiv
    \begin{pmatrix}
        0&\cdots&0&0\\
        \vdots&\ddots &\vdots&\vdots\\
        0&\cdots&|A_x(0)|^2&A_x(0)A^*_y(0)\\
        0&\cdots&A_y(0)A_x^*(0)&|A_y(0)|^2
    \end{pmatrix}
    =
    \frac{1}{2}
    \begin{pmatrix}
        0&\cdots&0&0\\
        \vdots&\ddots &\vdots&\vdots\\
        0&\cdots&I_0+Q_0&U_0-iV_0\\
        0&\cdots&U_0+iV_0&I_0-Q_0\\
         \end{pmatrix}
         =
    \begin{pmatrix}
        0&\cdots&0\\
        \vdots&\ddots &\vdots\\
        0&\cdots&\rho_{\gamma0}\\
         \end{pmatrix}
         ~.
    \label{eqrho0}
\end{align}
Note that when the initial photon state is fully polarized, the Stokes parameters satisfy $I_0^2=Q_0^2+U_0^2+V_0^2$, while in general they satisfy $I_0^2\geq Q_0^2+U_0^2+V_0^2\geq0$.

We consider the density matrix of the total system $\rho$, whose evolution along the propagation direction is described by the von Neumann-type equation
\begin{align}
    i\partial_z \rho=[\mathcal{M}(z),\rho]~,
    \label{eq:von}
\end{align}
where $\mathcal{M}$ denotes the total Hamiltonian.
It is convenient to separate the Hamiltonian into an unperturbed part $\mathcal{M}_0$ and a perturbation $\mathcal{M}_1(z)$:
\begin{align}
    \mathcal{M}(z)=\mathcal{M}_0+\mathcal{M}_1(z)~.
\end{align}
Here, $\mathcal{M}_0\equiv\mathrm{diag}[\Pi_{\varphi_1},\cdots,\Pi_{\varphi_N},\Pi_{\gamma},\Pi_{\gamma}]$ is the $z$-independent diagonal (unperturbed) Hamiltonian, whose components encode the dispersion relations of the corresponding fields in the absence of mixing.
The perturbation $\mathcal{M}_1(z)$, on the other hand, is generally $z$-dependent and describes the mixing between the photon sector and the other fields, while we assume that it has no entries in the $2\times 2$ photon sector.

Free propagation is described by the zeroth-order evolution operator,
\begin{align}
    \mathcal{U}^{(0)}(z) \equiv e^{-i\mathcal{M}_0z}~.
\end{align}
Introducing the interaction-picture Hamiltonian,
\begin{align}
    \mathcal{M}_I (z)
    \equiv \mathcal{U}^{(0)\dagger} (z) \, 
    \mathcal{M}_1 (z) \,
    \mathcal{U}^{(0)} (z)~,
\end{align}
the evolution of the total density matrix is then given by
\begin{align}
    \rho (z)
    =\mathcal{U}^{(0)} (z) \, [ \, \mathcal{U} (z)  \, \rho_0 \, \mathcal{U}^\dagger (z) \, ] \, \mathcal{U}^{(0)\dagger}(z) ~,
    \qquad
    \mathcal{U} (z)
    \equiv\mathrm{P}\exp \left[-i\int^z_0dz_1~\mathcal{M}_I(z_1) \right]~,
    \label{eq:propagation}
\end{align}
where $\mathrm{P}$ denotes the path-ordered product, and $\mathcal{U}(z)$ contains contributions to all orders in the perturbation.

\subsection{Born approximation}
\label{subsec:born}

In the weak-coupling regime, we can expand the unitary evolution operator $\mathcal{U}(z)$ perturbatively as
\begin{align}
\mathcal{U} (z)
= I_{(N+2) \times (N+2)} + \mathcal{U}^{(1)} (z) + \mathcal{U}^{(2)} (z) +\cdots~, 
\end{align}
where $I_{(N+2) \times (N+2)}$ denotes the $(N+2) \times (N+2)$ identity matrix, and the superscript $(i)$ of $\mathcal{U}^{(i)}$ denotes the order in perturbation theory.
The first- and second-order terms are given by
\begin{align}
    \mathcal{U}^{(1)} (z)
    \equiv-i\int_0^z dz_1~\mathcal{M}_I(z_1)~,\qquad 
    \mathcal{U}^{(2)} (z) 
    \equiv(-i)^2\int_0^zdz_1\int_0^{z_1}dz_2~\mathcal{M}_I(z_1)\mathcal{M}_I(z_2)~,
    \label{eq:2ndorder}
\end{align}
respectively.

Let us now focus on the Stokes parameters after propagation in the presence of photon-particle mixing.
We denote the two photon components by the labels $x$ and $y$, corresponding to the $(N+1)$-th and $(N+2)$-th components of the total density matrix $\rho$, respectively.
Using this notation, the  evolved Stokes parameters are obtained from the components of $\rho$ as
\begin{align}
I = \rho_{xx}+\rho_{yy}~,\qquad 
U = \rho_{xy}+\rho_{yx}~,\qquad 
V = i(\rho_{xy}-\rho_{yx})~,\qquad
Q = \rho_{xx}-\rho_{yy}~.
\end{align}
The total density matrix after propagation can be obtained by expanding Eq.~\eqref{eq:propagation} perturbatively.

In this work, since we assume that $\mathcal{M}_1$ has no entries in the $2 \times 2$ photon sector, $\mathcal{U}^{(1)}$ has no components within that sector.
Thus, the leading nontrivial evolution of the photon density matrix arises at second order when an initial state contains only photons, as in Eq.~\eqref{eqrho0}.
The evolution of the photon sector can be written as
\begin{align}
    \delta\rho_\gamma=U^{(0)}U^{(2)}\rho_{\gamma0}U^{(0)\dagger}+U^{(0)}\rho_{\gamma0}U^{(2)\dagger}U^{(0)\dagger}=U^{(2)}\rho_{\gamma0}+\rho_{\gamma0}U^{(2)\dagger}~,
    \label{eq:deltarho}
\end{align}
where $U^{(0)}$ and $U^{(2)}$ are the $2\times2$ photon-sector blocks of the evolution matrices of the total system $\mathcal{U}^{(0)}$ and $\mathcal{U}^{(2)}$, respectively.
In the last equality, we have used $ U^{(0)}=e^{-i\Pi_\gamma z}I_{2\times2}$, which commutes with the photon-sector matrices and whose overall phase cancels between $U^{(0)}$ and $U^{(0)\dagger}$.

Recalling the general $GL(2,\mathbb{C})$ transformation discussed in Sec.~\ref{subsec:EvolutionStokesParameters}, the second-order evolution operator $U^{(2)}$ can be identified with the infinitesimal generator introduced in Eq.~\eqref{eq:algebra} and decomposed into its Hermitian and anti-Hermitian parts as
\begin{align}
    U^{(2)}=-iX=-\frac12(k_\mu+ij_\mu)\sigma_\mu~.
\end{align}
Substituting $U^{(2)}$ into Eq.~\eqref{eq:deltarho} and comparing the resulting expression with Eq.~\eqref{eq:matrix}, we can express the eight transformation parameters $k_\mu$ and $j_\mu$ in terms of the evolution operator $U^{(2)}$ as
\begin{align}
k_0&= -\mathrm{Re}[U_{xx}^{(2)}+U_{yy}^{(2)}]~,
\qquad
k_1+ik_2= -(U_{yx}^{(2)}+U_{xy}^{(2)*})~,
\qquad 
k_3= -\mathrm{Re}[U_{xx}^{(2)}-U_{yy}^{(2)}]~,
\notag\\
j_0&= -\mathrm{Im}[U_{xx}^{(2)}+U_{yy}^{(2)}]~,
\qquad
j_1+ij_2 = i(U_{yx}^{(2)}-U_{xy}^{(2)*})~,
\qquad 
j_3= -\mathrm{Im}[U_{xx}^{(2)}+U_{yy}^{(2)*}]~.
\label{eq:defpara}
\end{align}
Note that the evolution of the total system is unitary, and hence the trace of the total density matrix is conserved, $\mathrm{Tr}[\rho]=\mathrm{Tr}[\rho_{\gamma0}]$, which implies $\sum_{n=1}^N \rho_{nn}+I=I_0$.
Accordingly, at second order in the perturbation, the transformation parameters $k_\mu$ can also be expressed in terms of bilinear combinations of the first-order evolution operator $\mathcal{U}^{(1)}$ as
\begin{align}
    k_0=\frac12\sum_{n=1}^N (|\mathcal{U}_{nx}^{(1)}|^2+|\mathcal{U}_{ny}^{(1)}|^2)~,\qquad
    k_1+ik_2=\sum_{n=1}^N\mathcal{U}_{nx}^{(1)}\mathcal{U}_{ny}^{(1)*}~,\qquad
    k_3=\frac12\sum_{n=1}^N (|\mathcal{U}_{nx}^{(1)}|^2-|\mathcal{U}_{ny}^{(1)}|^2 )~.
\end{align}

Equation~\eqref{eq:matrix} can also be written as the following infinitesimal transformation among the Stokes parameters,
\begin{align}
    \begin{pmatrix}
        \delta I\\\delta U\\\delta V\\\delta Q
    \end{pmatrix}
    &=-
    \begin{pmatrix}
        k_0&k_1&k_2&k_3  \\
        k_1&k_0&j_3&-j_2\\
        k_2&-j_3&k_0&j_1\\
        k_3&j_2&-j_1&k_0
        \end{pmatrix}
    \begin{pmatrix}
            I_0\\U_0\\V_0\\Q_0
        \end{pmatrix}
        ~.
        \label{eq:Stokes}
\end{align}
The overall phase parameter $j_0$ does not appear in this expression since it cancels out in the density matrix and therefore has no effect on the Stokes parameters.

\subsection{Kramers--Kronig relations between transformation parameters}
\label{subsec:KK}

The transformation parameters $k_\mu$ and $j_\mu$ introduced above are not independent.
As we show below, the causal structure of the second-order evolution operator $\mathcal{U}^{(2)}$ leads to Kramers--Kronig (KK) relations between them.
To see this, we rewrite the second-order evolution operator in Eq.~\eqref{eq:2ndorder} as
\begin{align}
    \mathcal{U}^{(2)}(z)=(-i)^2\int_0^z dz_1 \int_0^z dz_2~\theta(z_1-z_2)\mathcal{M}_I(z_1)\mathcal{M}_I(z_2)~,
\label{eq:2ndorderd2}
\end{align}
The step function originates from the path ordering and therefore encodes the causal ordering of the two interactions.

We now focus on the contribution to the photon block of the second-order evolution operator $U^{(2)}$ arising from the conversion into an additional degree of freedom $\varphi$.
Since we focus on the analytic structure with respect to the momentum difference $\Pi=\Pi_{\varphi\gamma}\equiv\Pi_{\varphi}-\Pi_\gamma$, we explicitly indicate the $\Pi$-dependence of the relevant quantities, regarding them as functions of $\Pi$.
Since $\Pi_\varphi$ and $\Pi_\gamma$ enter Eq.~(\ref{eq:2ndorderd2}) only through the diagonal Hamiltonian $\mathcal{M}_0$ and not through the interaction matrix $\mathcal{M}_1$, the phase part $e^{-i\Pi\tau}$ can be extracted from the photon sector of the product ${\mathcal M}_I (z_1) {\mathcal M}_I (z_2)$.
Thus the second-order evolution operator for the photon sector can be written as
\begin{align}
    U^{(2)}(\Pi)=-\frac12\left[k_\mu (\Pi)+i j_{\mu}(\Pi)\right]\sigma_\mu=-\int_0^z dz_1 \int_0^z dz_2~\theta(\tau) \, e^{-i\Pi \tau}\,T(z_1,z_2)~,
    \quad\tau\equiv z_1-z_2~,
    \label{eq:2ndphoton}
\end{align}
where $T(z_1,z_2)$ is a $\Pi$-independent $2\times2$ matrix.

When multiple fields $\varphi_i~(i=1,\cdots,N)$ are present, Eq.~\eqref{eq:2ndphoton} is understood as the contribution from each conversion channel with  $\Pi_{\varphi_i\gamma}\equiv\Pi_{\varphi_i}-\Pi_\gamma$.
The full second-order photon evolution is obtained by summing over all additional degrees of freedom, $U^{(2)}=\sum_{i=1}^N U_i^{(2)}(\Pi_{\varphi_i\gamma})$.
For simplicity, in the following discussion, we focus on the contribution from a single conversion channel.
In the graviton-photon system discussed later, the two graviton polarization states have the same dispersion relation and hence the same $\Pi_h$, so that their contributions can simply be combined by an overall factor of two.

Using the Fourier representation of the step function, the factor $\theta(\tau)e^{-i\Pi \tau}$ can be expressed as
\begin{align}
    \theta(\tau)e^{-i\Pi\tau}=\frac{i}{2\pi}\int_{-\infty}^{\infty}d\Omega~\frac{e^{-i\Omega\tau}}{\Omega-\Pi+i0}~.
\label{eq:reterded}
\end{align}
Thus, the second-order evolution operator has the retarded analytic structure associated with the denominator $(\Omega-\Pi+i0)^{-1}$.
Equivalently, using the following formula,
\begin{align}
    \frac{1}{\Omega-\Pi+i0}={\rm P.V.}\frac{1}{\Omega-\Pi}-i\pi\delta(\Omega-\Pi)~,
\end{align}
where ${\rm P.V.}$ denotes the Cauchy principal value, Eq.~\eqref{eq:2ndphoton} can be rewritten as 
\begin{align}
    U^{(2)}(\Pi)&=-\frac12\int_0^z dz_1 \int_0^z dz_2~\Big[e^{-i\Pi\tau}+\frac{i}{\pi}\mathrm{P.V.}\int_{-\infty}^{\infty}d\Omega \frac{e^{-i\Omega\tau }}{\Omega-\Pi}\Big]~T(z_1,z_2)\notag\\
    &=-\frac12[f(\Pi)+ig(\Pi)]~,
    \label{eq:fg1}
\end{align}
where 
\begin{align}
    f(\Pi)\equiv\int_0^z dz_1 \int_0^z dz_2~e^{-i\Pi\tau}\,T(z_1,z_2)~,\qquad g(\Pi)\equiv \frac{1}{\pi} \mathrm{P.V.}\int_{-\infty}^{\infty}d\Omega~\frac{f(\Omega)}{\Omega-\Pi}~.
    \label{eq:fg2}
\end{align}
Introducing the Hilbert transform of a function $h(\Pi)$ as
\begin{align}
    \mathcal{H}[h](\Pi)\equiv\frac{1}{\pi}\,\mathrm{P.V.}\int_{-\infty}^{\infty}d\Omega~\frac{h(\Omega)}{\Omega-\Pi}~,
    \label{eq:Hilbert}
\end{align}
we find from Eq.~\eqref{eq:fg2} that $f(\Pi)$ and $g(\Pi)$ are not independent but form a pair under the Hilbert transform as a consequence of the retarded structure of the second-order evolution,
\begin{align}
g(\Pi)=\mathcal{H}[f](\Pi)~,\qquad f(\Pi)=-\mathcal{H}[g](\Pi)~,
\label{eq:Hilbert_pair}
\end{align}
where we have used $\mathcal{H}^2=-1$.
We call this relationship the KK relation.
The KK pair also satisfies
\begin{align}
\int_{-\infty}^{+\infty} d\Pi~|g(\Pi)|^2
&=
\int_{-\infty}^{+\infty} d\Pi~|f(\Pi)|^2~.
\label{eq:Hilbert_pair_int}
\end{align}

The Hilbert transformation of $g(\Pi)$ can be explicitly performed using
\begin{align}
    \mathcal{H}\left[e^{-i\Pi\tau}\right]=-i~\mathrm{sgn}(\tau)e^{-i\Pi\tau}~,
\end{align}
which gives
\begin{align}
g(\Pi)=-i\int_0^z dz_1 \int_0^z dz_2~\mathrm{sgn}(\tau)e^{-i\Pi\tau}\,T(z_1,z_2)~.
\end{align}
Equation~\eqref{eq:fg1} therefore takes an equivalent form
\begin{align}
    U^{(2)}(\Pi)=-\frac12\int_0^z dz_1 \int_0^z dz_2~\Big[1+\mathrm{sgn}(\tau)\Big]e^{-i\Pi\tau}\,T(z_1,z_2)~.
    \label{eq:U2}
\end{align}
Since the interaction Hamiltonian of the total system is Hermitian, $T(z_1,z_2)$ satisfies $T^\dagger(z_1,z_2)=T(z_2,z_1)$.
Using this property and exchanging the integration variables $z_1\leftrightarrow z_2$, we find
\begin{align}
    U^{(2)\dagger}(\Pi)=-\frac12\int_0^z dz_1 \int_0^z dz_2~\Big[1+\mathrm{sgn}(\tau)\Big]e^{+i\Pi\tau}\,T^\dagger(z_1,z_2)=-\frac12\int_0^z dz_1 \int_0^z dz_2~\Big[1-\mathrm{sgn}(\tau)\Big]e^{-i\Pi\tau}\,T(z_1,z_2)~.
\end{align}
It then follows that the Hermitian and anti-Hermitian parts of $U^{(2)}$ are respectively given by
\begin{align}
    \frac{U^{(2)}+U^{(2)\dagger}}{2}(\Pi)&=-\frac12\int_0^z dz_1 \int_0^z dz_2~e^{-i\Pi\tau}\,T(z_1,z_2)=-\frac12 f(\Pi)~,\\
    \frac{U^{(2)}-U^{(2)\dagger}}{2i}(\Pi)&=\frac{i}{2}\int_0^z dz_1 \int_0^z dz_2~\mathrm{sgn}(\tau)e^{-i\Pi\tau}\,T(z_1,z_2)=-\frac12 g(\Pi)~.
\end{align}
On the other hand, these parts can also be expressed using the transformation parameters as
\begin{align}
    \frac{U^{(2)}+U^{(2)\dagger}}{2}(\Pi)&=-\frac12k_\mu(\Pi)\sigma_\mu~,\\
    \frac{U^{(2)}-U^{(2)\dagger}}{2i}(\Pi)&=-\frac12j_\mu(\Pi)\sigma_\mu~,
\end{align}
giving
\begin{align}
    k_\mu(\Pi)\sigma_\mu&=f(\Pi)=-\mathcal{H}[g](\Pi)=-\mathcal{H}[j_\mu](\Pi)\sigma_\mu~,\\
    j_\mu(\Pi)\sigma_\mu&=g(\Pi)=\mathcal{H}[f](\Pi)=\mathcal{H}[k_\mu](\Pi)\sigma_\mu~.
\end{align}
Therefore, the KK relations between $f$ and $g$ translate into
\begin{align}
    k_\mu(\Pi)=-\frac{1}{\pi}{\rm P.V.}\int_{-\infty}^{\infty}d\Omega~\frac{j_\mu(\Omega)}{\Omega-\Pi}~,
\qquad
j_\mu(\Pi)=\frac{1}{\pi}{\rm P.V.}\int_{-\infty}^{\infty}d\Omega~\frac{k_\mu(\Omega)}{\Omega-\Pi}~.
\label{eqKKkj}
\end{align}
Thus, the transformation parameters form KK pairs as a consequence of the causal structure encoded in the path-ordered second-order evolution.
These relations show that the non-unitary and unitary parts of the effective photon evolution are not independent: the parameters $k_\mu$, which describe attenuation associated with photon conversion, are related by causality to the parameters $j_\mu$, which describe the evolution of the polarization preserving the photon intensity.
Note that, although $j_0$ does not contribute to the evolution of the Stokes parameters as shown in Eq.~\eqref{eq:Stokes}, it remains a component of the effective photon evolution and forms a KK pair with the overall attenuation parameter $k_0$.
Also note that, when either $k_\mu$ or $j_\mu$ vanishes identically, so does its KK partner.

As we will see explicitly in the following sections, the KK relations have a direct physical consequence for photon conversion in background magnetic fields: whenever photon conversion induces a nontrivial change in circular polarization, the corresponding KK partner necessarily gives rise to birefringence, and vice versa.

\section{Conversion in a background magnetic field}
\label{sec:magnetic}

We now specialize the general formalism to photon conversion into other particles in the presence of a background magnetic field.
The background magnetic field enables the propagating photon to mix with additional degrees of freedom with spin $0$ or $2$.
In this section, we consider three representative systems: scalar-photon, pseudoscalar-photon, and tensor-photon conversion, corresponding to dilaton, axion, and graviton interactions, respectively.
Then, we identify the eight transformation parameters in each system that characterize the evolution of the Stokes parameters.

In the following, we consider photon propagation over a finite distance $d$.
We define the integration domains
\begin{align}
D \equiv [0,d]\times[0,d]~,\qquad
D_-\equiv \{(z_1,z_2)\in D~|~z_1\geq z_2\}~,
\end{align}
and introduce the following notation
\begin{align}
\int_D \equiv \int_0^d dz_1\int_0^d dz_2~,\qquad
\int_{D_-}\equiv \int_0^d dz_1\int_0^{z_1}dz_2~.
\end{align}
Equivalently, using $\tau= z_1-z_2$, the latter can be written as
\begin{align}
\int_{D_-}=\int_D\theta(\tau)=\frac{1}{2}\int_D\left[1+\mathrm{sgn}(\tau)\right]~.
\end{align}
We also introduce the following quadratic combinations of the magnetic-field components,
\begin{align}
C^\pm(z_1,z_2)&\equiv B_x(z_1)B_x(z_2)\pm B_y(z_1)B_y(z_2)=C^\pm(z_2,z_1)~, \notag \\
S^\pm(z_1,z_2)&\equiv B_x(z_1)B_y(z_2)\pm B_y(z_1)B_x(z_2) =\pm S^\pm(z_2,z_1)  ~.
\label{eqCS}
\end{align}

\subsection{Photon-to-dilaton conversion}

We first consider the dilaton-photon system in a background magnetic field.
The dilaton $\phi$ is taken to be a parity-even scalar field with a light mass $m_\phi$, coupled to the electromagnetic field through the interaction
\begin{align}
\mathcal{L}_{\phi\gamma}
=-\frac{g_{\phi\gamma\gamma}}{4}\phi F_{\mu\nu}F^{\mu\nu}
=-\frac{g_{\phi\gamma\gamma}}{2}\phi(\mathbf{B}^2-\mathbf{E}^2)~,
\end{align}
where $g_{\phi\gamma\gamma}$ denotes the dilaton-photon coupling constant.

To describe the dilaton-photon mixing, we decompose the electromagnetic field into a background magnetic field $\bar{\mathbf{B}}=(B_x,B_y,B_z)$ and a propagating photon field, $\mathbf{E}=-\partial_t\mathbf{A},~\mathbf{B}=\bar{\mathbf{B}}+\nabla\times \mathbf{A}$, where the photon is assumed to propagate along the $z$-direction in the radiation gauge \eqref{eq:gauge}.
Keeping only terms up to quadratic order in the propagating fields $(\phi,\mathbf{A})$, the interaction becomes
\begin{align}
\mathcal{L}_{\phi\gamma}
=-g_{\phi\gamma\gamma}\phi [B_yA_x^\prime-B_xA_y^\prime]~.
\end{align}
We decompose the dilaton and photon fields into monochromatic modes with angular frequency $\omega$ as
\begin{align}
     \phi(t,z)=\phi(z)e^{-i\omega t}+ \text{c.c.}~,\qquad A_m (t,z)= iA_m(z) \, e^{-i\omega t} + \text{c.c.}~.
\end{align}
Using the relativistic approximation, $\partial^2  \approx 2 \omega(\omega + i\partial_z) $, the equations of motion derived from the above Lagrangian reduce to the Schr\"{o}dinger-like propagation equation
\begin{align}
    i\partial_z\Psi_{\phi\gamma}(z)=[\mathcal{M}_{\phi\gamma0}+\mathcal{M}_{\phi\gamma1}(z)]\Psi_{\phi\gamma}(z)~.
\end{align}
Here, the field vector $\Psi_{\phi\gamma}$ and the unperturbed and mixing matrices are given by
\begin{align}
\Psi_{\phi\gamma}\equiv
\begin{pmatrix}
\phi\\A_x\\A_y    
\end{pmatrix}~,
\quad
  \mathcal{M}_{\phi\gamma0}\equiv
  \begin{pmatrix}
  \Pi_\phi(\omega)&0&0\\ 0&\Pi_\gamma(\omega)&0\\  0&0&\Pi_\gamma(\omega)
\end{pmatrix}~,
\quad
\mathcal{M}_{\phi\gamma1}(z)\equiv -\frac{g_{\phi\gamma\gamma}}{2}
  \begin{pmatrix}
  0&B_y(z)&-B_x(z)\\ B_y(z)&0&0\\  -B_x(z)&0&0
\end{pmatrix}~,
\end{align}
where $\Pi_\phi(\omega)\equiv-\omega+ m_\phi^2 / (2\omega)$ encodes the dispersion relation of the dilaton with mass $m_\phi$.

Using the mixing matrix $\mathcal{M}_{\phi\gamma1}$, the second-order evolution operator $U^{(2)}_{\phi\gamma}$ acting on the $2\times2$ photon sector is given by
\begin{align}
  U_{\phi\gamma}^{(2)}(d)=-c_{\phi\gamma}\int_{D_-}~e^{-i\Pi_{\phi\gamma} \tau}
  \begin{pmatrix}
  B_y(z_1)B_y(z_2)&-B_y(z_1)B_x(z_2)\\-B_x(z_1)B_y(z_2)&B_x(z_1)B_x(z_2)
\end{pmatrix}~,\qquad\tau=z_1-z_2~,
\end{align}
where $c_{\phi\gamma}\equiv g^2_{\phi\gamma\gamma}/4$ and $\Pi_{\phi\gamma}=\Pi_{\phi\gamma}(\omega)\equiv\Pi_\phi(\omega)-\Pi_\gamma(\omega)$.
Substituting $U^{(2)}_{\phi\gamma}$ into Eq.~\eqref{eq:defpara}, we obtain the transformation parameters $k_\mu$ and $j_\mu$ for photon-to-dilaton conversion as
\begin{align}
\begin{cases}
    k_{\phi0}(\omega)= \displaystyle c_{\phi\gamma} \int _{D_-} \cos[\Pi_{\phi\gamma}\tau]C^+(z_1,z_2)~,\\[0.3cm]
    k_{\phi1}(\omega)= \displaystyle-c_{\phi\gamma} \int _{D_-} \cos[\Pi_{\phi\gamma}\tau]S^+(z_1,z_2)~,\\[0.3cm]
    k_{\phi2}(\omega)= \displaystyle c_{\phi\gamma} \int _{D_-} \sin[\Pi_{\phi\gamma}\tau]S^-(z_1,z_2)~,\\[0.3cm]
    k_{\phi3}(\omega)= \displaystyle-c_{\phi\gamma} \int _{D_-} \cos[\Pi_{\phi\gamma}\tau]C^-(z_1,z_2) ~,\\[0.3cm]
    \end{cases}
    ~,\qquad 
    \begin{cases}
 j_{\phi0}(\omega)= -
 \displaystyle c_{\phi\gamma}\int_{D_-} \sin[\Pi_{\phi\gamma}\tau]C^+(z_1,z_2)~,\\[0.3cm]
    j_{\phi1}(\omega)= \displaystyle c_{\phi\gamma}\int_{D_-} \sin[\Pi_{\phi\gamma}\tau]S^+(z_1,z_2) ~,\\[0.3cm]
    j_{\phi2}(\omega)= \displaystyle c_{\phi\gamma}\int_{D_-} \cos[\Pi_{\phi\gamma}\tau] S^-(z_1,z_2) ~,\\[0.3cm]
    j_{\phi3}(\omega)= \displaystyle c_{\phi\gamma}\int_{D_-} \sin[\Pi_{\phi\gamma}\tau]C^-(z_1,z_2)~.
    \end{cases}
    \label{eq:dparas}
\end{align}
Considering Eq.~\eqref{eq:U2}, the above relations can be rewritten as
\begin{align}
\begin{cases}
    k_{\phi0}(\omega)&= \displaystyle \frac12 c_{\phi\gamma} \int _{D} e^{-i\Pi_{\phi\gamma}\tau}C^+(z_1,z_2)~,\\[0.3cm]
    k_{\phi1}(\omega)&= \displaystyle-\frac12c_{\phi\gamma} \int _{D} e^{-i\Pi_{\phi\gamma}\tau}S^+(z_1,z_2)~,\\[0.3cm]
    k_{\phi2}(\omega)&= \displaystyle \frac i2c_{\phi\gamma} \int _{D} e^{-i\Pi_{\phi\gamma}\tau}S^-(z_1,z_2)~,\\[0.3cm]
    k_{\phi3}(\omega)&= \displaystyle-\frac12c_{\phi\gamma} \int _{D} e^{-i\Pi_{\phi\gamma}\tau}C^-(z_1,z_2) ~,\\[0.3cm]
    \end{cases}
    ~,\qquad 
    \begin{cases}
 j_{\phi0}(\omega)&= 
 \displaystyle -\frac i2c_{\phi\gamma}\int_{D} \mathrm{sgn}(\tau)e^{-i\Pi_{\phi\gamma}\tau}C^+(z_1,z_2)~,\\[0.3cm]
    j_{\phi1}(\omega)&= \displaystyle \frac i2 c_{\phi\gamma}\int_{D} \mathrm{sgn}(\tau)e^{-i\Pi_{\phi\gamma}\tau}S^+(z_1,z_2) ~,\\[0.3cm]
    j_{\phi2}(\omega)&= \displaystyle \frac12 c_{\phi\gamma}\int_{D} \mathrm{sgn}(\tau)e^{-i\Pi_{\phi\gamma}\tau} S^-(z_1,z_2) ~,\\[0.3cm]
    j_{\phi3}(\omega)&= \displaystyle \frac i2c_{\phi\gamma}\int_{D} \mathrm{sgn}(\tau)e^{-i\Pi_{\phi\gamma}\tau}C^-(z_1,z_2)~.
    \end{cases}
\end{align}
From the above expressions, one can verify the following identity,
\begin{align}
    k_{\phi0}^2=k_{\phi1}^2+k_{\phi2}^2+k_{\phi3}^2~.
    \label{eq:iddilaton}
\end{align}
Using Eq.~(\ref{eq:Hilbert_pair_int}), and noting that $j_{\phi \mu}$ and $k_{\phi \mu}$ are real-valued functions, Eq.~(\ref{eq:iddilaton}) also implies
\begin{align}
\int_{-\infty}^{+\infty}d\Pi~
j_{\phi 0}^2(\Pi)
&=
\int_{-\infty}^{+\infty}d\Pi~
\sum_{i = 1}^3 j_{\phi i}^2(\Pi)~.
\label{eq:iddilaton_int}
\end{align}

\subsection{Photon-to-axion conversion}

Next, we consider the axion-photon system.
The axion $a$ is taken to be a parity-odd pseudoscalar field with a light mass $m_a$, coupled to the electromagnetic field through
\begin{align}
\mathcal{L}_{a\gamma}
=-\frac{g_{a\gamma\gamma}}{4}a F_{\mu\nu}\tilde{F}^{\mu\nu}
=-g_{a\gamma\gamma}a\mathbf{B}\cdot\mathbf{E}~,
\end{align}
where $g_{a\gamma\gamma}$ denotes the axion-photon coupling constant and $\tilde{F}^{\mu\nu}=(1/2)\epsilon^{\mu\nu\rho\sigma}F_{\rho\sigma}$ is the dual electromagnetic field strength tensor.

Following the same procedure as in the dilaton case, we decompose the electromagnetic field into a background magnetic field and a propagating photon field.
The quadratic interaction terms reduce to
\begin{align}
    \mathcal{L}_{a\gamma}
=g_{a\gamma\gamma}a[B_x\dot{A}_x+B_y\dot{A}_y]~,
\end{align}
and we can obtain the equations of motion for the axion-photon system from this Lagrangian.
We decompose the axion and photon fields into monochromatic modes with angular frequency $\omega$ as
\begin{align}
      a(t,z)=a(z)e^{-i\omega t}+ \text{c.c.}~,\qquad A_m (t,z)= iA_m(z) \, e^{-i\omega t} + \text{c.c.}~.
\end{align}
Using the relativistic approximation, the equations of motion reduce to the Schr\"{o}dinger-like equation,
\begin{align}
    i\partial_z\Psi_{a\gamma}(z)=[\mathcal{M}_{a\gamma0}+\mathcal{M}_{a\gamma1}(z)]\Psi_{a\gamma}(z)~.
\end{align}
Here, the field vector $\Psi_{a\gamma}$ and the unperturbed and mixing matrices are given by
\begin{align}
\Psi_{a\gamma}\equiv
\begin{pmatrix}
    a\\A_x\\A_y
\end{pmatrix}
~,
\quad
\mathcal{M}_{a\gamma0}=
  \begin{pmatrix}
  \Pi_a(\omega)&0&0\\ 0&\Pi_\gamma(\omega)&0\\ 0&0&\Pi_\gamma(\omega)
\end{pmatrix}~,
\quad
\mathcal{M}_{a\gamma1}(z)= -\frac {g_{a\gamma\gamma}}{2}
  \begin{pmatrix}
  0&B_x(z)&B_y(z)\\ ~B_x(z)&0&0\\ ~B_y(z)&0&0
\end{pmatrix}~,
\end{align}
where $\Pi_a(\omega)\equiv-\omega+m_a^2/(2\omega)$ encodes the dispersion relation of the axion with a light mass $m_a$.

From the above mixing matrix, the second-order perturbative evolution operator $U_{a\gamma}^{(2)}$ acting on the $2\times2$ photon sector is given by
\begin{align}
 U_{a\gamma}^{(2)}(d)=-c_{a\gamma}\int_{D_-} ~e^{-i\Pi_{a\gamma} \tau}
  \begin{pmatrix}
B_x(z_1)B_x(z_2)&B_x(z_1)B_y(z_2)\\B_y(z_1)B_x(z_2)&B_y(z_1)B_y(z_2)
\end{pmatrix}
~,\qquad \tau=z_1-z_2
~,
\end{align}
where $c_{a\gamma}\equiv g^2_{a\gamma\gamma}/4$ and $\Pi_{a\gamma}=\Pi_{a\gamma}(\omega)\equiv\Pi_a(\omega)-\Pi_\gamma(\omega)$.
Substituting $U_{a\gamma}^{(2)}$ into Eq.~\eqref{eq:defpara}, we obtain the transformation parameters $k_\mu$ and $j_\mu$ for photon-to-axion conversion (the Primakoff effect) as
\begin{align}
\begin{cases}
    k_{a0}= \displaystyle c_{a\gamma}\int_{D_-} \cos[\Pi_{a\gamma}\tau]C^+(z_1,z_2) ~,\\[0.3cm]
    k_{a1}= \displaystyle c_{a\gamma}\int_{D_-} \cos[\Pi_{a\gamma}\tau]S^+(z_1,z_2)~,\\[0.3cm]
    k_{a2}= \displaystyle c_{a\gamma}\int_{D_-} \sin[\Pi_{a\gamma}\tau]S^-(z_1,z_2) ~,\\[0.3cm]
    k_{a3}= \displaystyle c_{a\gamma}\int_{D_-} \cos[\Pi_{a\gamma}\tau]C^-(z_1,z_2) ~.
    \end{cases}
    ~,\qquad
    \begin{cases}
    j_{a0}= \displaystyle -c_{a\gamma}\int_{D_-} \sin[\Pi_{a\gamma}\tau]C^+(z_1,z_2) ~,\\[0.3cm]
    j_{a1}= \displaystyle -c_{a\gamma}\int_{D_-} \sin[\Pi_{a\gamma}\tau]S^+(z_1,z_2) ~,\\[0.3cm]
    j_{a2}= \displaystyle c_{a\gamma}\int_{D_-} \cos[\Pi_{a\gamma}\tau]S^-(z_1,z_2) ~,\\[0.3cm]
    j_{a3}= \displaystyle -c_{a\gamma}\int_{D_-} \sin[\Pi_{a\gamma}\tau]C^-(z_1,z_2) ~.
    \end{cases}
    \label{eq:aparas}
\end{align}
As in the dilaton-photon system, using the relation in Eq.~\eqref{eq:U2}, we can rewrite the above expressions as
\begin{align}
\begin{cases}
    k_{a0}&= \displaystyle \frac12 c_{a\gamma}\int_{D}e^{-i\Pi_{a\gamma}\tau}C^+(z_1,z_2) ~,\\[0.3cm]
    k_{a1}&= \displaystyle \frac12  c_{a\gamma}\int_{D} e^{-i\Pi_{a\gamma}\tau}S^+(z_1,z_2) ~,\\[0.3cm]
    k_{a2}&=\displaystyle \frac i2  c_{a\gamma}\int_{D} e^{-i\Pi_{a\gamma}\tau}S^-(z_1,z_2) ~,\\[0.3cm]
    k_{a3}&= \displaystyle \frac12  c_{a\gamma}\int_{D} e^{-i\Pi_{a\gamma}\tau}C^-(z_1,z_2) ~.
    \end{cases}
    ~,\qquad
    \begin{cases}
    j_{a0}&= \displaystyle -\frac i2 c_{a\gamma}\int_{D} \mathrm{sgn}(\tau)e^{-i\Pi_{a\gamma}\tau}C^+(z_1,z_2) ~,\\[0.3cm]
    j_{a1}&= \displaystyle -\frac i2c_{a\gamma}\int_{D} \mathrm{sgn}(\tau)e^{-i\Pi_{a\gamma}\tau}S^+(z_1,z_2) ~,\\[0.3cm]
    j_{a2}&= \displaystyle \frac12 c_{a\gamma}\int_{D} \mathrm{sgn}(\tau)e^{-i\Pi_{a\gamma}\tau}S^-(z_1,z_2) ~,\\[0.3cm]
    j_{a3}&= \displaystyle -\frac i2c_{a\gamma}\int_{D} \mathrm{sgn}(\tau)e^{-i\Pi_{a\gamma}\tau}C^-(z_1,z_2) ~.
    \end{cases}
\end{align}
Comparing the dilaton-photon and axion-photon systems, we find that the parameters $k_0$, $k_2$, $j_0$, and $j_2$ have identical structures, whereas the parameters $k_1$, $k_3$, $j_1$, and $j_3$ differ only by their relative signs.
Thus, these parameters also satisfy 
\begin{align}
    k_{a0}^2=k_{a1}^2+k_{a2}^2+k_{a3}^2~.
    \label{eq:idaxion}
\end{align}
As in the dilaton case, using Eq.~(\ref{eq:Hilbert_pair_int}) and noting that $j_{a \mu}$ and $k_{a \mu}$ are real-valued functions, Eq.~(\ref{eq:idaxion}) also implies
\begin{align}
\int_{-\infty}^{+\infty}d\Pi~
j_{a 0}^2(\Pi)
&=
\int_{-\infty}^{+\infty}d\Pi~
\sum_{i = 1}^3 j_{a i}^2(\Pi)~.
\label{eq:idaxion_int}
\end{align}

\subsection{Photon-to-graviton conversion}

Finally, we consider the graviton-photon system.
In linearized general relativity, the spacetime metric $g_{\mu\nu}$ is expanded around the Minkowski metric $\eta_{\mu\nu}$ as
\begin{align}
    g_{\mu\nu}=\eta_{\mu\nu}+\frac{2}{M_{\mathrm{P}}}h_{\mu\nu}~,
\end{align}
where $h_{\mu\nu}$ denotes the graviton field and $M_{\rm P} = 2.4\times10^{18}\,\mathrm{GeV}$ is the reduced Planck mass.
We impose the transverse-traceless gauge condition on the graviton field,
\begin{align}
    h_{0\mu}=0~,\qquad\partial_jh_{ij}=0~,\qquad h_{ii}=0~.
\end{align}
The graviton field couples to the energy-momentum tensor of the electromagnetic field through the linear interaction as
\begin{align}
\mathcal{L}_{h\gamma}
=\frac{1}{M_{\mathrm{P}}}h_{\mu\nu}T_{\mathrm{EM}}^{\mu\nu}~,
\qquad
T_{\mathrm{EM}}^{\mu\nu}\equiv F^{\mu\alpha}F^{\nu}_{~~\alpha}-\frac{1}{4}g^{\mu\nu}F_{\alpha\beta}F^{\alpha\beta}~.
\end{align}

Unlike the above two cases, the graviton possesses two independent polarization modes.
We therefore introduce a polarization basis for a graviton propagating along the $z$ direction and expand the graviton field in terms of the polarization tensors $\varepsilon^{(n)}_{ij}(n=+,\times)$ as
\begin{align}
    h_{ij}(t,\bm{x})\equiv\sum_{n=+,\times} h_n(t,z)\varepsilon_{ij}^{(n)}~,
\end{align}
where
\begin{align}
    \varepsilon_{ij}^{(+)}
    \equiv \frac{1}{\sqrt{2}} (e_i^{(x)}e_j^{(x)}-e_i^{(y)} e_j^{(y)})
    =\frac{1}{\sqrt{2}}
    \begin{pmatrix}
        1&0&0\\
        0&-1&0\\
        0&0&0\\
    \end{pmatrix}
    ~,\qquad
    \varepsilon_{ij}^{(\times)}
    \equiv \frac{1}{\sqrt{2}} (e_i^{(x)}e_j^{(y)}+e_i^{(y)}e_j^{(x)})
    =\frac{1}{\sqrt{2}}
    \begin{pmatrix}
        0&1&0\\
        1&0&0\\
        0&0&0\\
    \end{pmatrix}
    ~.
\end{align}

Using the polarization expansion above and decomposing the electromagnetic field into a background magnetic field and a propagating photon field, the quadratic interaction Lagrangian is separated into contributions from the two graviton polarization modes as
\begin{align}
    \mathcal{L}_{h_+\gamma}
=\frac{\sqrt{2}}{M_{\mathrm{P}}}h_+[B_xA^\prime_y+B_yA^\prime_x]~,
\qquad \mathcal{L}_{h_\times\gamma}
=\frac{\sqrt{2}}{M_{\mathrm{P}}}h_\times[B_yA^\prime_y-B_xA^\prime_x]~.
\end{align}
The equations of motion for the graviton-photon system can be derived from this Lagrangian.
We decompose the graviton and photon fields into monochromatic modes with angular frequency $\omega$ as
\begin{align}
    h_{n}(t,z)=h_n(z)\,e^{-i\omega t} + \text{c.c.}
    ~,\qquad A_m (t,z)= iA_m(z) \, e^{-i\omega t} + \text{c.c.}~.
\end{align}
Using the relativistic approximation, we obtain the propagation equation as
\begin{align}
    i\partial_z\Psi_{h\gamma}(z)=[\mathcal{M}_{h\gamma0}+\mathcal{M}_{h\gamma1}(z)]\Psi_{h\gamma}(z)~.
\end{align}
Here, we define
\begin{align}
\Psi_{h\gamma}\equiv
\begin{pmatrix}
h_+\\h_\times\\A_x\\A_y    
\end{pmatrix}
~,
\quad
\mathcal{M}_{h\gamma0}=
  \begin{pmatrix}
  \Pi_h&0&0&0\\ 0&\Pi_h&0&0\\0&0&\Pi_\gamma&0\\ 0&0&0&\Pi_\gamma
\end{pmatrix}~,
\quad
\mathcal{M}_{h\gamma1}(z)= \frac {1}{\sqrt{2}M_\mathrm{P}}
  \begin{pmatrix}
  0&0&B_y(z)&B_x(z)\\ 0&0&-B_x(z)&B_y(z)\\B_y(z)&-B_x(z)&0&0\\ B_x(z)&B_y(z)&0&0
\end{pmatrix}~,
\end{align}
where $\Pi_{h_+}=\Pi_{h_\times}=\Pi_h(\omega)\equiv-\omega$, corresponding to the common dispersion relation for the two polarization modes of a massless graviton.

Using the mixing matrix $\mathcal{M}_{h\gamma1}$, the second-order perturbative evolution operator in the photon sector is given by
\begin{align}
U_{h\gamma}^{(2)}(d)=-c_{h\gamma}\int_{D_-} ~e^{-i\Pi_{h\gamma} \tau}
\begin{pmatrix}
    C^+(z_1,z_2)&-S^-(z_1,z_2)\\S^-(z_1,z_2)&C^+(z_1,z_2)
\end{pmatrix}
~,\qquad\tau=z_1-z_2~,
\end{align}
where $c_{h\gamma}\equiv1/(2M_\mathrm{P}^2)$ and $\Pi_{h\gamma}=\Pi_{h\gamma}(\omega)\equiv \Pi_h-\Pi_\gamma$.
Note that, for the graviton-photon system, $\Pi_{h\gamma}$ vanishes in vacuum since both particles are massless.
However, in realistic astrophysical environments, the photon acquires an effective plasma mass $m_{\rm{pl}}$ due to the finite electron number density $n_e$, given by
\begin{align}
    m_\mathrm{pl}=\sqrt{\frac{4\pi \alpha n_e}{m_e}} ~,
\end{align}
with $\alpha\approx1/137$ the fine-structure constant and $m_e$ the electron mass.
This effective mass gives rise to a nonzero  $\Pi_{h\gamma}(\omega)\simeq-m_{\rm pl}^2/(2\omega)$.
We retain $\Pi_{h\gamma}$ to keep the expressions in a general form.

Substituting $U_{h\gamma}^{(2)}$ into Eq.~\eqref{eq:defpara}, we obtain the transformation parameters $k_\mu$ and $j_\mu$ for photon-to-graviton conversion (the Gertsenshtein effect) as
\begin{align}
\begin{cases}
    k_{h0}= \displaystyle 2 c_{h\gamma}\int_{D_-} \cos[\Pi_{h\gamma}\tau] C^+(z_1,z_2) ~,\\[0.3cm]
    k_{h1}= \displaystyle0~,\\[0.3cm]
    k_{h2}= \displaystyle -2 c_{h\gamma}\int_{D_-} \sin[\Pi_{h\gamma}\tau] S^-(z_1,z_2)~,\\[0.3cm]
    k_{h3}= \displaystyle0~,
    \end{cases}
    ~,~
    \begin{cases}
 j_{h0}= \displaystyle  -2 c_{h\gamma}\int_{D_-} \sin[\Pi_{h\gamma}\tau] C^+(z_1,z_2)~,\\[0.3cm]
    j_{h1}= 0~,\\[0.3cm]
    j_{h2}= \displaystyle  -2c_{h\gamma}\int_{D_-} \cos[\Pi_{h\gamma}\tau] S^-(z_1,z_2)~,\\[0.3cm]
     j_{h3}= 0~.
\end{cases}
\label{eq:gparas}
\end{align}
Using the relation in Eq.~\eqref{eq:U2}, these expressions can be rewritten as
\begin{align}
\begin{cases}
    k_{h0}&= \displaystyle  c_{h\gamma}\int_{D} e^{-i\Pi_{h\gamma}\tau} C^+(z_1,z_2) ~,\\[0.3cm]
    k_{h1}&= \displaystyle0~,\\[0.3cm]
    k_{h2}&= \displaystyle -i c_{h\gamma}\int_{D}  e^{-i\Pi_{h\gamma}\tau} S^-(z_1,z_2)~,\\[0.3cm]
    k_{h3}&= \displaystyle0~,
    \end{cases}
    ~,~
    \begin{cases}
 j_{h0}&= \displaystyle  -i c_{h\gamma}\int_{D} \mathrm{sgn}(\tau) e^{-i\Pi_{h\gamma}\tau} C^+(z_1,z_2)~,\\[0.3cm]
    j_{h1}&= 0~,\\[0.3cm]
    j_{h2}&= \displaystyle  -c_{h\gamma}\int_{D}  \mathrm{sgn}(\tau) e^{-i\Pi_{h\gamma}\tau}S^-(z_1,z_2)~,\\[0.3cm]
     j_{h3}&= 0~.
\end{cases}
\end{align}

In the graviton-photon system, the parameters $k_1$, $k_3$, $j_1$, and $j_3$ vanish identically because conversion into the two graviton polarizations, $+$ and $\times$, contributes equally to the evolution of linear polarization.
The remaining nonvanishing parameters $k_0$, $k_2$, $j_0$, and $j_2$ differ from the corresponding parameters in the dilaton- and axion-photon systems by an overall factor of two.
Furthermore, $k_2$ and $j_2$ have opposite signs compared with those in the dilaton- and axion-photon systems.
In contrast to the dilaton and axion cases, the corresponding parameters in the graviton-photon system do not satisfy the identity given in Eqs.~\eqref{eq:iddilaton} and~\eqref{eq:idaxion}.

\section{Conversion in stochastic magnetic fields}
\label{sec:random}

In this section, we consider the case in which the background magnetic field is stochastic and characterized by its power spectra.
In Sec.~\ref{subsec:all}, we represent the formal solution of the ensemble-averaged total density matrix over the stochastic magnetic field, including contributions to all orders in the interaction.
In Sec.~\ref{subsec:average}, we express the expectation values of the photon Stokes parameters in terms of the magnetic-field power spectra.
 
We assume a statistically homogeneous, isotropic, and Gaussian magnetic field, whose statistical properties are fully described by the two-point correlation function,
\begin{align}
  \left< B_i(\vec{x})B_j(\vec{x}')\right>\equiv \int_ke^{i\vec{k}\cdot(\vec{x}-\vec{x}')}\Big[(\delta_{ij}-\hat{k}_i\hat{k}_j )P_B(k)-i\epsilon_{ijm}\hat{k}_mP_{aB}(k)\Big]~,\qquad 
  \int_k\equiv\int \frac{d^3k}{(2\pi)^3}~,
  \label{2ptfunc}
\end{align}
where $\hat{k}_i\equiv k_i/k$ with $k\equiv\sqrt{k_x^2+k_y^2+k_z^2}$.
The bracket $\left< \cdots \right>$ denotes an ensemble average over infinitely many realizations of the stochastic magnetic field.
$P_B(k)$ and $P_{aB}(k)$ denote the symmetric and antisymmetric parts of the magnetic-field power spectrum, respectively.
The combination $(\delta_{ij} - \hat{k}_i \hat{k}_j)$ in Eq.~\eqref{2ptfunc} reflects the divergence-free nature of the magnetic field.
The factor $i$ in the second term is included so that $P_{aB}(k)$ is real, as is $P_B(k)$.
When $P_{aB}(k) \neq 0$, the magnetic field is referred to as helical.
Maximal helicity is realized when $|P_{aB}(k)| = P_B(k)$.

\subsection{All-order ensemble-averaged evolution in Gaussian stochastic fields}
\label{subsec:all}

Instead of working directly with Eq.~\eqref{eq:von}, let us consider the von Neumann equation in the interaction picture,
\begin{align}
    i\partial_z \rho_I = [\mathcal{M}_I(z),\rho_I]\equiv\mathcal{L}_{\mathcal{M}_I(z)} \rho_I~,
\end{align}
where we define the Liouvillian-like operator by $\mathcal{L}_X(\cdot)\equiv[X,\cdot]$ and the interaction-picture density matrix by  $\rho_I\equiv\mathcal{U}^{(0)\dagger}\rho\,\mathcal{U}^{(0)}$.
Using this notation, the formal solution is given by
\begin{align}
    \rho_I(z)=\mathrm{P}\exp \left[ -i\int^z_0dz_1\mathcal{L}_{\mathcal{M}_I(z_1)} \right] \rho_{I}(0)~.
    \label{eq:formalsol}
\end{align}

We assume that the interaction-picture Hamiltonian $\mathcal{M}_I$ depends linearly on the stochastic variables $B_i(z)$ as
\begin{align}
    \mathcal{M}_I(z)=\sum_{m=x,y} B_m(z)T_m(z)\equiv B_m(z)T_m(z)~,
\end{align}
where $T_m(z)$ denotes the $z$-dependent matrix  associated with the magnetic-field component $B_m (z)$, and summation over repeated indices $m = x,y$ is understood.
All the photon-conversion models  discussed above have a Hamiltonian of this form.
Therefore, the following formulation can be directly applied to these photon-conversion processes.
Using the above notation, the operator $\mathcal{L}_{\mathcal{M}_I}$ can be factorized into the stochastic variable and the matrix part as $\mathcal{L}_{\mathcal{M}_I(s)}=B_m(s)\mathcal{L}_{T_m(s)}$.
The formal solution \eqref{eq:formalsol} can then be expanded as
\begin{align}
    \rho_I(d)&=\mathrm{P}\exp \left[-i\int^d_0ds~\mathcal{L}_{\mathcal{M}_I(s)} \right]\rho_{I}(0)\nonumber\\
    &=\mathrm{P}\sum_{n=0}^\infty \frac{(-i)^n}{n!}\int_0^d dz_1\cdots \int_0^d dz_n~B_{m_1}(z_1)\cdots B_{m_n}(z_n)\,\mathcal{L}_{T_{m_1}(z_1)}\cdots \mathcal{L}_{T_{m_n}(z_n)}\,\rho_{I}(0)~.
    \label{eq:rhoI}
\end{align}

Because of the Gaussian nature of $B_m(z)$, all higher-order correlation functions can be reduced to sums of products of two-point correlation functions according to Wick's theorem.
From this property, upon ensemble averaging Eq.~\eqref{eq:rhoI}, the odd-order terms vanish, while the even-order terms are reduced to products of two-point correlation functions.
Thus, only the terms with $n=2k$ contribute, where $k$ is a nonnegative integer.
The $2k$-point correlation function can be decomposed into a sum of $(2k-1)!!$ pairwise contractions, so that
\begin{align}
    \left<\rho_I(d)\right>&= \mathrm{P}\sum_{k=0}^\infty \frac{(-i)^{2k}}{2k!}\int_0^d dz_1\int_0^d dz_2\cdots \int_0^d dz_{2k}\nonumber\\
    &\quad\times[\left<  B_{m_1}(z_1)B_{m_2}(z_2)\right>\cdots\left< B_{m_{2k-1}}(z_{2k-1})B_{m_{2k}}(z_{2k})\right>+(\text{other combinations})]\nonumber\\
    &\quad\times\mathcal{L}_{T_{m_1}(z_1)}\mathcal{L}_{T_{m_2}(z_2)}\cdots \mathcal{L}_{T_{m_{2k}}(z_{2k})}\,\rho_{I}(0)\nonumber\\
    &=\mathrm{P}\sum_{k=0}^\infty \frac{(-i)^{2k}(2k-1)!!}{2k!}\int_0^d dz_1\cdots \int_0^d dz_{2k}\nonumber\\
    &\quad\times\left<  B_{m_1}(z_1)B_{m_2}(z_2)\right>\cdots\left< B_{m_{2k-1}}(z_{2k-1})B_{m_{2k}}(z_{2k})\right>\mathcal{L}_{T_{m_1}(z_1)}\cdots \mathcal{L}_{T_{m_{2k}}(z_{2k})}\,\rho_{I}(0)~.
\end{align}
In the last step, we used the fact that the path-ordered product is invariant under the simultaneous interchange of the integration variables and their associated operators, $\mathrm{P}[\mathcal{L}_{T_{m(z)}}\mathcal{L}_{T_{m'(z')}}]=\mathrm{P}[\mathcal{L}_{T_{m'(z')}}\mathcal{L}_{T_{m(z)}}]$, so that, after relabeling the dummy integration variables and summed indices, all Wick-contraction terms give identical contributions.
Finally, using the identity, 
\begin{align}
    (2k-1)!!=\frac{(2k)!}{(2)^kk!}~,
\end{align}
we obtain the following all-order expression,
\begin{align}
    \left<\rho_I(d)\right>&= \mathrm{P}\sum_{k=0}^\infty \frac{1}{k!}\Big(-\frac{1}{2}\Big)^k\int_0^d dz_1\cdots \int_0^d dz_{2k}\nonumber\\
    &\quad\times\left< B_{m_1}(z_1)B_{m_2}(z_2)\right>\cdots\left< B_{m_{2k-1}}(z_{2k-1})B_{m_{2k}}(z_{2k})\right>\mathcal{L}_{T_{m_1}(z_1)}\cdots \mathcal{L}_{T_{m_{2k}}(z_{2k})}\,\rho_{I}(0)\nonumber\\
    &=\mathrm{P}\sum_{k=0}^\infty  \frac{1}{k!} \Big(-\frac12 \int_0^d ds\int_0^d ds'\left<  B_{m}(s)B_{m'}(s')\right> \mathcal{L}_{T_{m}(s)}\mathcal{L}_{T_{m'}(s')}\Big)^k\rho_{I}(0)\nonumber\\
    &=\mathrm{P}\exp \Big[-\frac12 \int_0^d ds\int_0^d ds'\left<  B_{m}(s)B_{m'}(s')\right> \mathcal{L}_{T_{m}(s)}\mathcal{L}_{T_{m'}(s')}\Big]\rho_{I}(0)\nonumber\\
    &=\mathrm{P}\exp \Big[-\frac12 \int_0^d ds\int_0^d ds'\left<  \mathcal{L}_{\mathcal{M}_I(s)}\mathcal{L}_{\mathcal{M}_I(s')}\right> \Big]\rho_{I}(0)~.
\end{align}
This result is essentially analogous to the relation $\left<e^{ix}\right>_x=e^{\left<-\frac12 x^2\right>_x}$, where $\left<\cdots\right>_x$ denotes the ensemble average with respect to a Gaussian random variable $x$.
Therefore, the ensemble-averaged evolution of the full system at arbitrary order in the stochastic field can be expressed in terms of its power spectra.

\subsection{Ensemble average of Stokes parameters}
\label{subsec:average}

As shown in Sec.~\ref{sec:magnetic}, all the parameters governing the Stokes-parameter transformation in each photon-particle system are quadratic in the magnetic-field components and can be expressed in terms of the four combinations $C^\pm (z_1,z_2)$ and $S^\pm (z_1,z_2)$ defined in Eq.~\eqref{eqCS}.
Assuming a statistically homogeneous and isotropic magnetic field, two of these combinations vanish upon ensemble averaging, $\left<C^- (z_1,z_2) \right>=\left<S^+ (z_1,z_2) \right>=0$.
Therefore, $\left< j_{1}\right>$, $\left< j_{3}\right>$, $\left< k_1 \right>$ and $\left<k_{3}\right>$ vanish in Eqs.~\eqref{eq:dparas} and \eqref{eq:aparas}, whereas $j_1$, $j_3$, $k_1$ and $k_{3}$ trivially vanish in the graviton-photon system.
In contrast, both $\left<C^+ (z_1,z_2) \right>$ and $\left<S^- (z_1,z_2) \right>$ generally remain nonzero, with the latter arising from magnetic helicity.
We therefore define the following four types of integrals involving the nonvanishing ensemble-averaged correlations with $\tau=z_1-z_2$:
\begin{align}
  \begin{cases}
  \alpha\equiv\displaystyle\int_{D_-} \cos[\Pi\tau] \langle{ C^+(z_1,z_2) \rangle}=\displaystyle\frac12\int_{D} e^{-i\Pi\tau} \langle{ C^+(z_1,z_2) \rangle}~,\\[0.3cm]
    \chi\equiv \displaystyle-\int_{D_-} \sin[\Pi\tau] \langle{ C^+(z_1,z_2) \rangle}=\displaystyle-\frac i2\int_{D}\mathrm{sgn}(\tau) e^{-i\Pi\tau} \langle{ C^+(z_1,z_2)\rangle}~,\\[0.3cm]
    \beta\equiv \displaystyle\int_{D_-} \sin[\Pi\tau] \langle{ S^-(z_1,z_2) \rangle}=\displaystyle \frac i2\int_{D} e^{-i\Pi\tau} \langle{ S^-(z_1,z_2) \rangle}~,\\[0.3cm]
    \psi\equiv \displaystyle\int_{D_-} \cos[\Pi\tau] \langle{ S^-(z_1,z_2) \rangle}=\displaystyle\frac12\int_{D} \mathrm{sgn}(\tau)e^{-i\Pi\tau} \langle{ S^-(z_1,z_2)\rangle}~,
    \end{cases}
  \label{defalphabetapsi}
\end{align}
These quantities are proportional to the ensemble-averaged transformation parameters $k_0, j_0, k_2,$ and $j_2$, respectively.
Note that $(k_0, j_0)$ and $(k_2, j_2)$ form KK pairs as seen in Eq.~\eqref{eqKKkj}, and consequently $(\alpha,\chi)$ and $(\beta,\psi)$ form KK pairs.

Combining the above discussion with Eq.~\eqref{eq:Stokes}, the ensemble-averaged Stokes parameters can be decomposed into two independent sectors; the intensity and circular-polarization sector $(I,V)$ and the linear-polarization sector $(Q,U)$.
Also, the expectation values of the evolved Stokes parameters can be written in terms of the above integrals.
For the dilaton-photon and axion-photon systems, the expectation values of the changes in the Stokes parameters are obtained as
\begin{align}
    \begin{pmatrix}
       \left< \delta I\right>\\\left<\delta V\right>
    \end{pmatrix}
    =-c_{\varphi\gamma}
    \begin{pmatrix}
        \alpha&\beta  \\
        \beta&\alpha
        \end{pmatrix}
        \begin{pmatrix}
            I_0\\V_0
        \end{pmatrix}
        ~,\qquad
    \begin{pmatrix}
       \left<\delta Q\right>\\\left<\delta U\right>
    \end{pmatrix}
    =-c_{\varphi\gamma}
    \begin{pmatrix}
        \alpha&\psi\\
        -\psi&\alpha
        \end{pmatrix}
        \begin{pmatrix}
           Q_0\\U_0
        \end{pmatrix}
        ~,
\end{align}
where $\varphi=\phi$ for the dilaton and $\varphi=a$ for the axion, and $c_{\varphi\gamma} \equiv g_{\varphi\gamma\gamma}^2/4$.
On the other hand, for the graviton-photon system, the expectation values of the changes in the Stokes parameters are given by
\begin{align}
    \begin{pmatrix}
       \left< \delta I\right>\\\left<\delta V\right>
    \end{pmatrix}
    =-2c_{h\gamma}
    \begin{pmatrix}
        \alpha&-\beta  \\
        -\beta&\alpha
        \end{pmatrix}
        \begin{pmatrix}
            I_0\\V_0
        \end{pmatrix}
        ~,\qquad
    \begin{pmatrix}
       \left<\delta Q\right>\\\left<\delta U\right>
    \end{pmatrix}
    =-2c_{h\gamma}
    \begin{pmatrix}
        \alpha&-\psi\\
        \psi&\alpha
        \end{pmatrix}
        \begin{pmatrix}
           Q_0\\U_0
        \end{pmatrix}
        ~,
\end{align}
where $c_{h\gamma} \equiv 1/(2M_{\mathrm{P}}^2)$.
Note that $\alpha$, $\chi$, $\beta$, and $\psi$ defined in Eq.~\eqref{defalphabetapsi} have mass dimension 2.
Combining them with either $c_{\varphi \gamma}$ or $c_{h\gamma}$, each of which has mass dimension $-2$, yields dimensionless coefficients multiplying the Stokes parameters.

Using the definition of the magnetic-field correlation function \eqref{2ptfunc}, the four integrals can be rewritten as the following convolutions:
\begin{align}
  \begin{cases}
  \alpha=\displaystyle\int _k~\mathcal{I}(k_z;\Pi)\frac{1+\hat{k}^2_z}{2}P_{B}(k)~,\\[0.3cm]
  \chi=\displaystyle\int _k~\mathcal{K}(k_z;\Pi)\frac{1+\hat{k}^2_z}{2}P_{B}(k)~,\\[0.3cm]
    \beta=\displaystyle\int_k~\mathcal{I}(k_z;\Pi)\hat{k}_z P_{aB}(k)~,\\[0.3cm]
    \psi= \displaystyle\int _k~\mathcal{K}(k_z;\Pi)\hat{k}_z P_{aB}(k)~.
  \end{cases}
  \label{eq:abpsi}
\end{align}
Here, the integral kernels are expressed as 
\begin{align}
  \mathcal
  {I}(\theta_-)
  &= d^2\left[ \frac{2 \sin (\theta_- /2)}{\theta_- } \right]^2=\mathcal{I}(-\theta_-)~,\\
  \mathcal{K}(\theta_-)
  &= d^2\Big[\frac{2\sin\theta_-}{\theta_-^2}-\frac{2}{\theta_-}\Big]=-\mathcal{K}(-\theta_-)~,
\end{align}
where
\begin{align}
    \theta_-\equiv\theta_-(k_z;\Pi(\omega))=(\Pi(\omega)- k_z)d~.
\end{align}
Note that the kernels $\mathcal{I}$ and $\mathcal{K}$ form a KK pair since $(\alpha, \chi)$ and $(\beta, \psi)$ each form a KK pair.

These typical integrals $\alpha,\chi,\beta,$ and $\psi$ can be reduced since the integrands in Eq.~\eqref{eq:abpsi} are rotationally invariant in the $x$-$y$ plane of the wavevector space.
For practical numerical computations, these integrals can thus be reduced to the following one-dimensional forms,
\begin{align}
  \begin{cases}
    \alpha(\Pi d)= \displaystyle \frac{1}{(2\pi)^2}\int_0^\infty dk~P_B(k) \, \mathcal{C}_{\alpha}(kd;\Pi d)~,\\[0.3cm]
    \chi(\Pi d)= \displaystyle \frac{1}{(2\pi)^2}\int_0^\infty dk~P_B(k) \, \mathcal{C}_{\chi}(kd;\Pi d)~,\\[0.3cm]
    \beta(\Pi d)= \displaystyle \frac{1}{(2\pi)^2} \int_0^\infty dk~P_{aB}(k) \, \mathcal{C}_{\beta}(kd;\Pi d)~,\\[0.3cm]
    \psi(\Pi d)= \displaystyle \frac{1}{(2\pi)^2}\int_0^\infty dk~P_{aB}(k) \, \mathcal{C}_{\psi}(kd;\Pi d)~.
  \end{cases}
  \label{eq:abxi}
\end{align}
Here, $\mathcal{C}_{i}$ $(i=\alpha,\chi, \beta, \psi)$ are the convolution kernels given by
\begin{align}
  \begin{cases}
       \mathcal{C}_{\alpha}(kd;\Pi d)\equiv\displaystyle(kd)^2\int_{-1}^1 d\mu~\left[ \frac{2\sin (\theta_-/2)}{\theta_-} \right]^2 \frac{1+\mu^2}{2}~,\\[0.3cm]
       \mathcal{C}_{\chi}(kd;\Pi d)\equiv\displaystyle(kd)^2\int_{-1}^1 d\mu~\left[\frac{2\sin\theta_-}{\theta_-^2}-\frac{2}{\theta_-}\right] \frac{1+\mu^2}{2}~,\\[0.3cm]
       \mathcal{C}_{\beta}(kd;\Pi d)\equiv \displaystyle(kd)^2\int_{-1}^1d\mu~\left[ \frac{2 \sin (\theta_-/2)}{\theta_-} \right]^2~\mu~,\\[0.3cm]
       \mathcal{C}_{\psi}(kd;\Pi d)\equiv\displaystyle(kd)^2\int_{-1}^{1}d\mu~\left[\frac{2\sin\theta_-}{\theta_-^2}-\frac{2}{\theta_-}\right]~\mu~,
    \end{cases}
    \label{eq:C}
\end{align}
with 
\begin{align}
    \theta_-=\Pi d- kd\cdot \mu~.
\end{align}
The integrals in Eq.~\eqref{eq:C} can be evaluated analytically, and the resulting expressions are summarized in Appendix \ref{app:A}.

The integrals $\alpha$ and $\beta$ also appeared in our previous studies of the axion-photon system \cite{Chiba:2025hgr} and the graviton-photon system \cite{Chiba:2025odu}, where we investigated the conversion probability and the circular polarization generated from initially unpolarized photons.
By extending the analysis to arbitrary initial polarization states, we can now clarify the physical roles of these integrals in the general evolution of the Stokes parameters.
In particular, $\alpha$, previously associated with the decrease in photon intensity due to conversion, is identified as a parameter that describes the common attenuation of all four Stokes parameters.
In addition, a new integral $\psi$ appears for initially polarized photons and describes the rotation of the linear-polarization plane, corresponding to birefringence.

The KK relations derived in Sec.~\ref{subsec:KK} further reveal that the effects described by the four integrals are not completely independent.
In particular, $\beta$, which describes the helicity-induced variation of circular polarization, and $\psi$, which describes birefringence, form a KK pair.
Therefore, a nontrivial contribution to the circular-polarization evolution is necessarily accompanied by a corresponding birefringent contribution, and vice versa.
The same analysis also reveals that $\alpha$ and $\chi$ form a KK pair.
Although the contribution of $\chi$ does not directly affect the Stokes parameters and was therefore absent in our previous analyses focused on photon observables, it constitutes the counterpart of the overall attenuation described by $\alpha$.
Thus, the present analysis reveals a common structure underlying attenuation, circular-polarization evolution, and birefringence in photon conversion into other particles, irrespective of whether photons convert into dilatons, axions, or gravitons.

\subsection{Probing magnetic helicity through photon polarization}

We now investigate how the magnetic helicity of the background field is encoded in the polarization evolution of photons. As shown in the previous subsection, the antisymmetric part of the magnetic power spectrum $P_{aB}$ gives rise to both circular polarization and birefringence, characterized by the integrals $\beta$ and $\psi$, respectively. Here, we demonstrate that these quantities provide a direct probe of the magnetic helicity density.

We introduce a vector potential $\bar{\mathbf A}=(A_x,A_y,A_z)$ for the background magnetic field, satisfying the Coulomb gauge condition $\nabla\cdot\bar{\mathbf A}=0$, such that
\begin{align}
    \mathbf{B}=(B_x,B_y,B_z)\equiv\nabla\times\bar{\mathbf{A}}~.
\end{align}
In Fourier space, this condition allows us to express the vector potential in terms of the magnetic field as
\begin{align}
    \tilde{\bm{A}}(\vec{k})=\frac{i}{k^2}\vec{k}\times\tilde{\bm{B}}(\vec{k})~,
\end{align}
where $\tilde{\mathbf A}$ and $\tilde{\mathbf B}$ denote the Fourier components of the vector potential and the magnetic field, respectively. Using the two-point correlation function in Eq.~\eqref{2ptfunc}, the ensemble-averaged magnetic helicity density is then given by
\begin{align}
    \langle h_M\rangle \equiv \langle \bar{\mathbf{A}}\cdot \mathbf{B}\rangle=\int_0^\infty dk \frac{k}{\pi^2}P_{aB}(k) ~.
\end{align}
Here, we have used the statistical homogeneity and isotropy of the magnetic field. Although the local magnetic helicity density $h_M=\bar{\mathbf A}\cdot\mathbf B$ is generally gauge dependent, its ensemble average can be expressed entirely in terms of the gauge-invariant magnetic-field correlation function.

We next consider $\psi$ defined in Eq.~\eqref{defalphabetapsi}. Using statistical homogeneity, the double integral can be reduced to
\begin{align}
    \psi(d;\Pi)=d\int_0^dd\tau \left(1-\frac{\tau}{d} \right)\cos(\Pi \tau)\langle{ S^-(z_1,z_2) \rangle}~,
\end{align}
where $\tau=z_1-z_2$.
From the magnetic-field correlation function in Eq.~\eqref{2ptfunc}, the antisymmetric correlation $S^-$, defined in Eq.~\eqref{eqCS}, can be expressed as
\begin{align}
    \langle{ S^-(z_1,z_2) \rangle}=2\int_{k}\sin(k_z\tau)\hat{k}_zP_{aB}(k)~.
\end{align}
Substituting this expression into the integral for $\psi$ and taking the limit $\Pi\to0$, we obtain
\begin{align}
    \frac{\psi(d;\Pi=0)}{d}=\int_0^\infty dk \frac{k}{\pi^2}P_{aB}(k)\Big(1-\frac{\mathrm{Si}(kd)}{kd}\Big)~.
\end{align}
When the propagation distance $d$ is much larger than the typical correlation length of the magnetic field, the integral approaches the expression for the magnetic helicity density. Consequently,
\begin{align}
  \langle h_M\rangle
=\lim_{d\to\infty}\frac{\psi(d;0)}{d}~.
\end{align}
Remarkably, since $\beta$ and $\psi$ form a Kramers–Kronig pair, the magnetic helicity density can also be reconstructed from the spectral dependence of the integral $\beta$:
\begin{align}
  \langle{ h_M\rangle}= \lim_{d\to\infty} \frac{\psi(d;0)}{d}=\lim_{d\to\infty}\frac{1}{\pi d}\mathrm{P.V.}
\int_{-\infty}^\infty d\Omega\frac{\beta(d;\Omega)}{\Omega}~.
\end{align}
This relation shows that the average magnetic helicity density can be extracted either from the birefringence accumulated per unit propagation distance or, equivalently, from the frequency-integrated circular polarization through the KK relation, independently of the specific particle species into which photons convert.
Note that, while birefringence requires initial linear polarization, circular polarization can be generated even from initially unpolarized photons. 
Therefore, the KK relation allows the magnetic helicity density to be reconstructed from the circular-polarization response even when birefringence is not directly observable.

\section{Numerical analysis}
\label{sec:numerical}

In this section, we first show the behavior of each convolution kernel as a function of $k d$.
We then adopt representative forms for $P_B (k)$ and $P_{aB} (k)$ and illustrate how the convolutions of the power spectra with these kernels behave as functions of $\Pi d$.
For convenience, throughout this section we write $\Pi_{\varphi\gamma}d$ simply as $\Pi d$.
For conversion into a massive particle of mass $m$, we have $\Pi d=m^2d/(2\omega)$, where $\omega$ is the angular frequency.
Thus, small values of $\Pi d$ correspond to the high-frequency regime, whereas large values correspond to the low-frequency regime.

\subsection{\texorpdfstring{Kernels $\mathcal{C}_\alpha$,  $\mathcal{C}_\chi$, $\mathcal{C}_\beta$, and $\mathcal{C}_\psi$}{Kernels Calpha, Cbeta, and Cpsi}}
\label{subsec:C}

\begin{figure}
\centering
\includegraphics[width=0.4\linewidth]{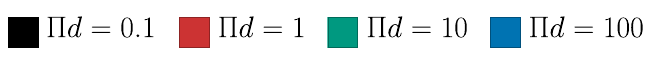}
\\[0.4cm]
\includegraphics[width=0.4\linewidth]{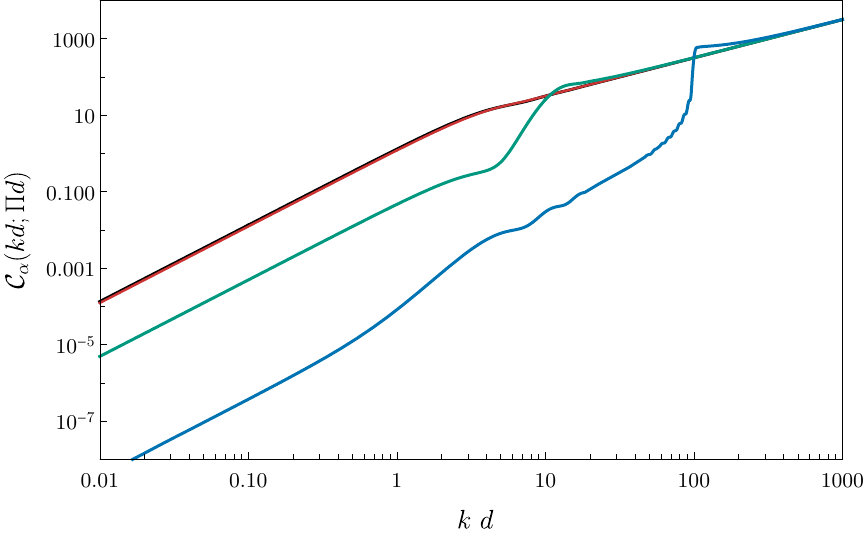}
\includegraphics[width=0.4\linewidth]{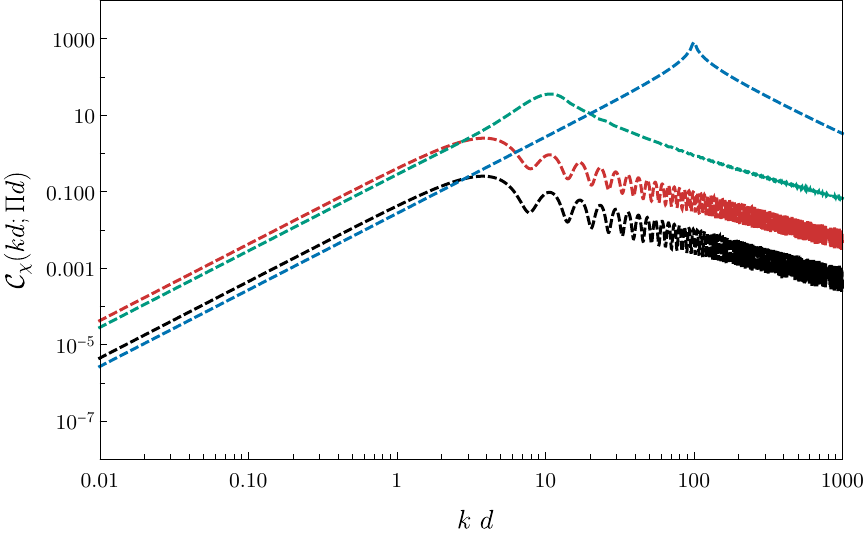}
\\[0.4cm]
\includegraphics[width=0.4\linewidth]{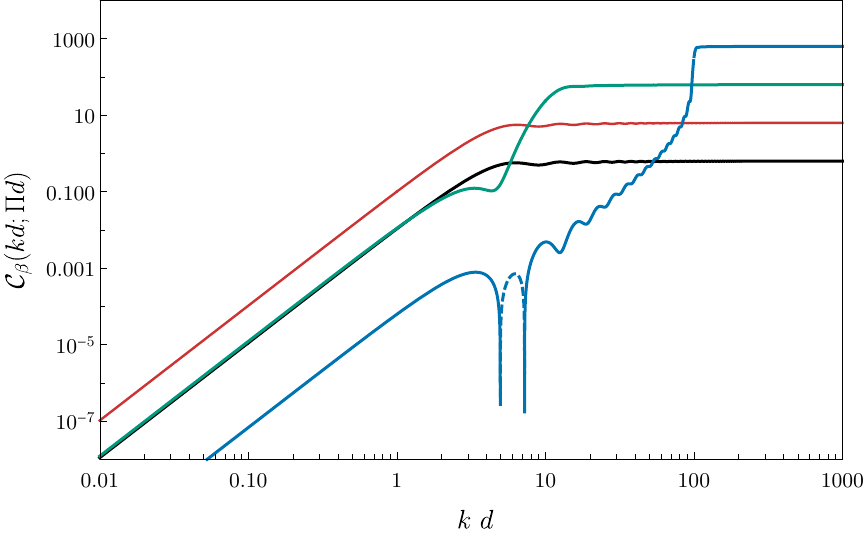}
\includegraphics[width=0.4\linewidth]{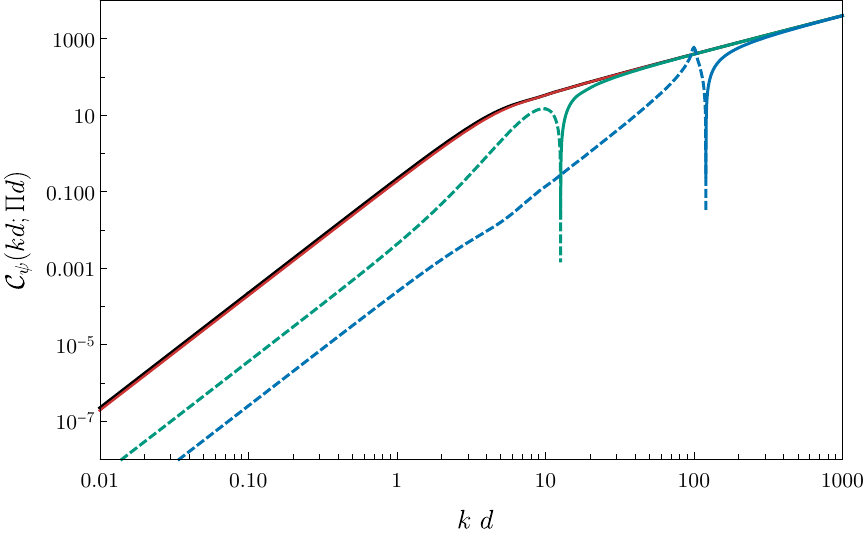}
\caption{
Convolution kernels $\mathcal{C}_\alpha$, $\mathcal{C}_\chi$, $\mathcal{C}_\beta$, and $\mathcal{C}_\psi$ as functions of $kd$ for various values of $\Pi d$.
The black, red, green, and blue curves correspond to $\Pi d = 0.1$, $1$, $10$, and $100$, respectively.
Dashed curves indicate regions where the kernels take negative values.
}
\label{fig:C}
\end{figure}

In Fig.~\ref{fig:C}, the integral kernels $\mathcal{C}_i~(i=\alpha,\chi,\beta,\psi)$ are shown as functions of the dimensionless parameter $kd$ for various values of $\Pi d$.
Therefore, the functional form of each kernel is independent of the magnetic-field properties and is common to all three systems by taking the ensemble average.

Let us highlight the main features of the kernels shown in Fig.~\ref{fig:C}.
Their behavior differs qualitatively depending on whether $\Pi d$ is smaller or larger than unity:
\begin{itemize}
\item 
$\Pi d \lesssim 1$: 
For small values of $\Pi d$, the kernels $\mathcal{C}_\alpha$ and $\mathcal{C}_\psi$ show almost no dependence on $\Pi d$, whereas both $|\mathcal{C}_\chi|$ and $\mathcal{C}_\beta$ scale linearly with $\Pi d$.
All four kernels exhibit a crossover around $kd=\mathcal{O}(1)$ between their infrared and ultraviolet scaling behaviors.
In the infrared regime $kd\lesssim1$, one finds $\mathcal{C}_\alpha, |\mathcal{C}_\chi|\propto(kd)^2$, while $\mathcal{C}_\beta$ and $\mathcal{C}_\psi$ scale as $(kd)^3$.
In the ultraviolet regime $kd\gtrsim1$, $\mathcal{C}_\alpha$ and $\mathcal{C}_\psi$ scale as $kd$, while $\mathcal{C}_\beta$ approaches a constant.
In contrast to the other kernels, $|\mathcal C_\chi|$ reaches a maximum at $kd=\mathcal O(1)$ and decreases toward large $kd$ while oscillating.

\item
$\Pi d \gg 1$: 
For large values of $\Pi d$, the qualitative behavior of the kernels changes around two characteristic scales, $kd=\mathcal{O}(1)$ and $kd\simeq\Pi d$.
For $kd\gtrsim\Pi d$, the absolute values of the kernels approach the same scaling behaviors as those found for $\Pi d\lesssim1$, whereas they are strongly suppressed for $kd\ll\Pi d$.
In the infrared regime $kd \lesssim 1$, the kernels behave as $\mathcal{C}_\alpha, |\mathcal{C}_\chi| \propto (kd)^2$ and $\mathcal{C}_\beta, |\mathcal{C}_{\psi}| \propto (kd)^3$.
Between the two characteristic scales, $1\lesssim kd\lesssim\Pi d$, the kernels exhibit nontrivial modulations.
The kernel $\mathcal{C}_\psi$ takes negative values for $kd\lesssim\Pi d$, and shows a sign change around $\Pi d\sim k d$.
In contrast to the other kernels, $|\mathcal C_\chi|$ develops a pronounced peak around $kd\simeq\Pi d$, and decreases for $kd\gtrsim\Pi d$.
\end{itemize}

\subsection{\texorpdfstring{Integrals $\alpha$, $\chi$, $\beta$, and $\psi$}{Integrals alpha, beta, and psi}}
\label{subsec:abx}

\begin{figure}[h!]
\centering
\includegraphics[width=0.4\linewidth]{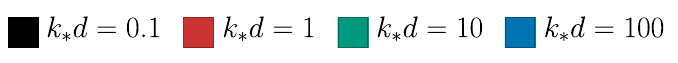}
\\
\includegraphics[width=0.38\linewidth]{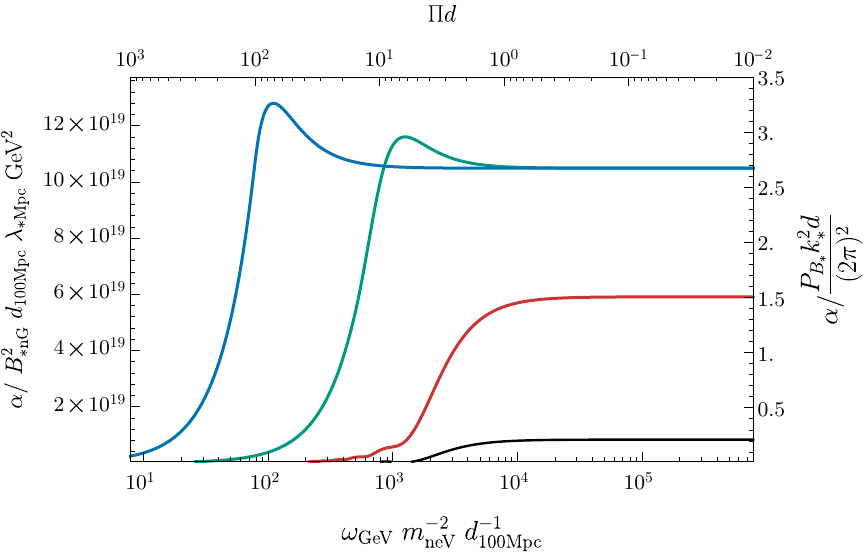}
\hskip 0.4cm
\includegraphics[width=0.38\linewidth]{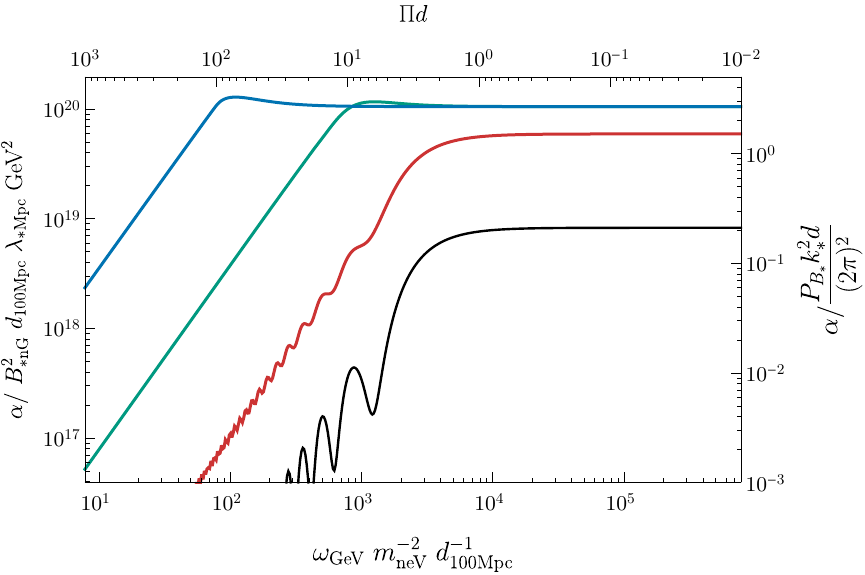}
\\
\includegraphics[width=0.38\linewidth]{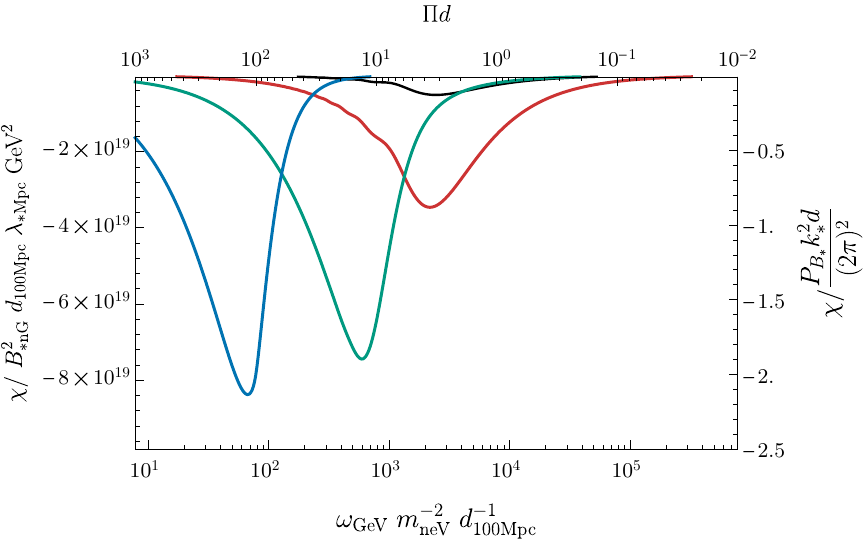}
\hskip 0.4cm
\includegraphics[width=0.38\linewidth]{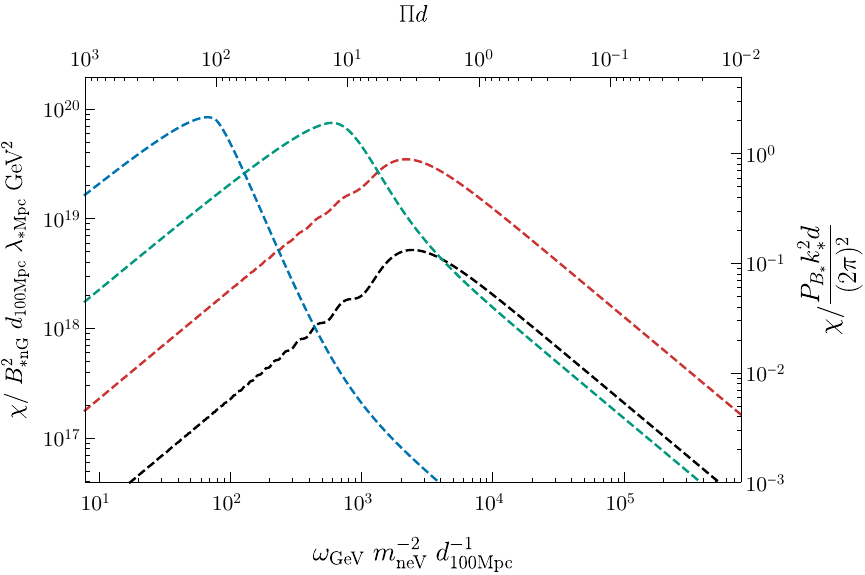}
\\
\includegraphics[width=0.38\linewidth]{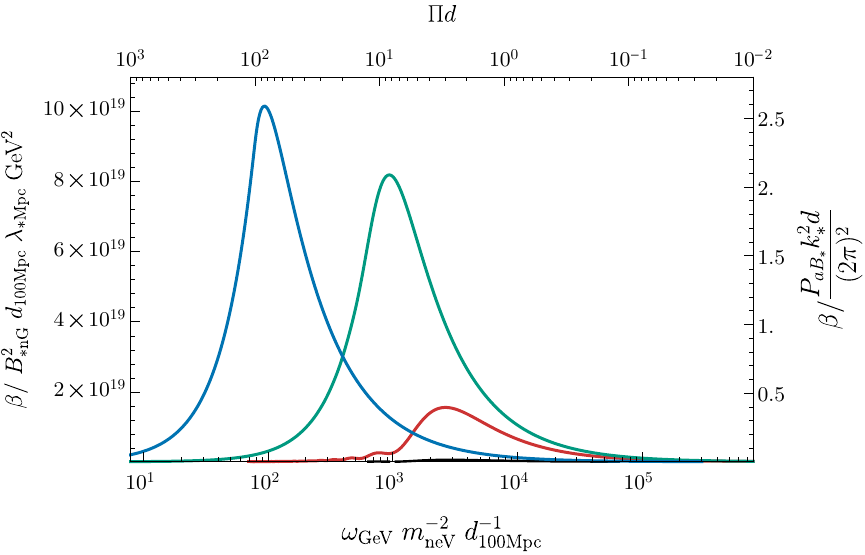}
\hskip 0.4cm
\includegraphics[width=0.38\linewidth]{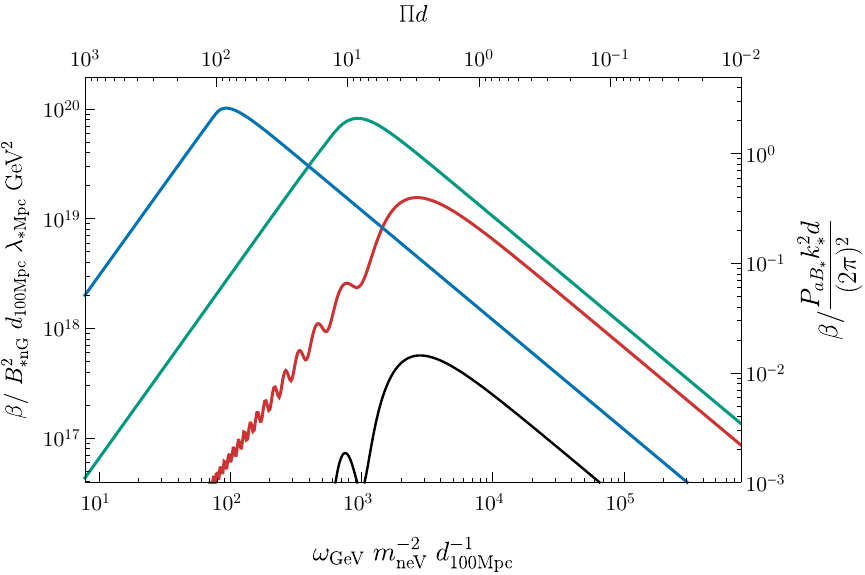}
\\
\includegraphics[width=0.38\linewidth]{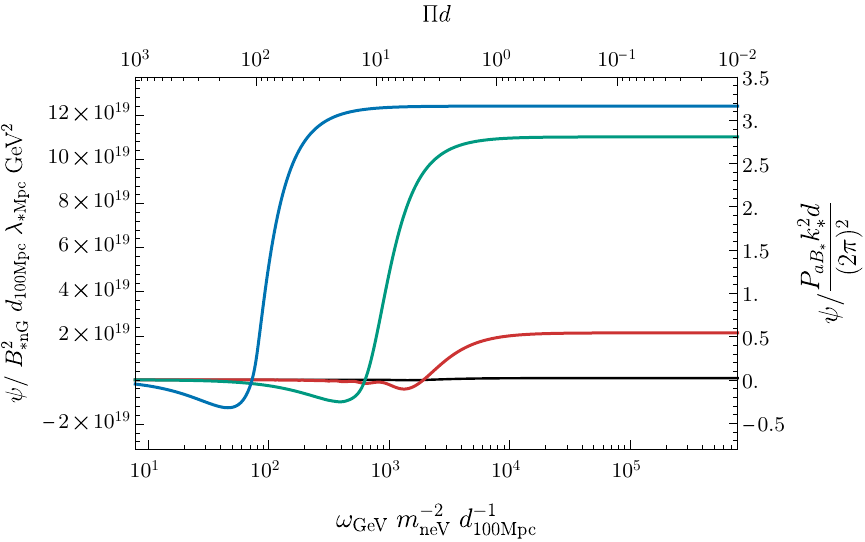}
\hskip 0.4cm
\includegraphics[width=0.38\linewidth]{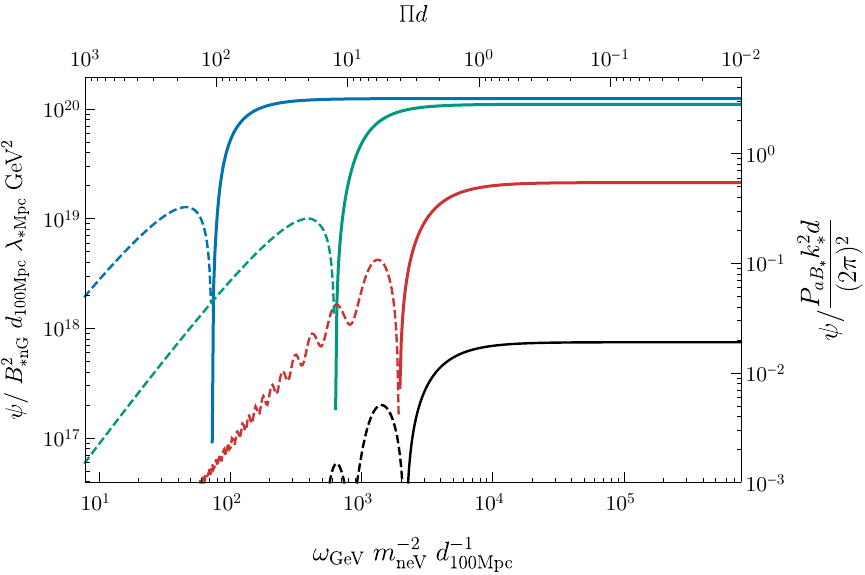}
\caption{
Integrals $\alpha$ (first row), $\chi$ (second row), $\beta$ (third row), and $\psi$ (fourth row), evaluated using the magnetic-field power spectra \eqref{eq:PB} and \eqref{eq:PaB}, as functions of the photon angular frequency $\omega$.
The black, red, green, and blue curves correspond to $k_* d = 0.1$, $1$, $10$, and $100$, respectively.
In each panel, the upper horizontal axis shows the dimensionless variable $\Pi d$, while the lower horizontal axis shows $\omega_{\mathrm{GeV}}\equiv\omega/\mathrm{GeV}$, normalized by $m_{\mathrm{neV}}\equiv m/\mathrm{neV}$ and $d_{100\mathrm{Mpc}}\equiv d/(100\mathrm{Mpc})$, where $m$ is the mass of the light particle and $d$ is the propagation distance.
The right vertical axis shows $\alpha$, $\chi$, $\beta$, or $\psi$, normalized by $P_{(a)B*}k_*^2d/(2\pi)^2$.
The left vertical axis shows the corresponding quantities normalized by $B_{*\mathrm{nG}}^2\equiv(B_*/\mathrm{nG})^2$, $d_{100\mathrm{Mpc}}$, $\lambda_{\mathrm{Mpc}}\equiv\lambda_*/\mathrm{Mpc}$, where $B_*$ is the magnetic-field strength at the peak scale $k_*$ and $\lambda_*\equiv2\pi/k_*$ is the corresponding characteristic length.
In each row, the left and right panels show the same quantity on linear and logarithmic vertical scales, respectively.
In the right panels of the second and fourth rows, dashed curves indicate regions where $\chi$ and $\psi$ take negative values.
}
\label{fig:abx}
\end{figure}

In Fig.~\ref{fig:abx}, the integrals $\alpha$, $\chi$, $\beta,$ and $\psi$ are shown as functions of the dimensionless parameter $\Pi d$, obtained by convolving each kernel with magnetic-field power spectra.
Here, we adopt the following functional forms of the power spectra presented in Ref.~\cite{Durrer:2013pga}:
\begin{align}
    P_B(k) &= P_{B*} \times
    \begin{cases}
    \displaystyle\left(\frac{k}{k_*}\right)^{n_l}~, \quad n_l = 2~, & (k \leq k_*) \\
    \displaystyle\left(\frac{k}{k_*}\right)^{n_h}~, \quad n_h = -\frac{11}{3}~, & (k > k_*) 
    \end{cases}
    \label{eq:PB}
    \\
   P_{aB}(k) &= P_{aB*} \times
    \begin{cases}
    \displaystyle\left(\frac{k}{k_*}\right)^{n_{al}}~, \quad n_{al} = 3~, & (k \leq k_*) \\
    \displaystyle\left(\frac{k}{k_*}\right)^{n_{ah}}~, \quad n_{ah} = - \frac{11}{3}~, & (k > k_*) 
    \end{cases}
    \label{eq:PaB}
\end{align}
where $P_{B*}$ and $P_{aB*}$ are the peak amplitudes of the symmetric and antisymmetric parts of the magnetic-field power spectrum, respectively (see Eq.~\eqref{2ptfunc}), and $k_*$ is the characteristic wavenumber of the magnetic field.
The exponents $n_l$ and $n_h$ ($n_{al}$ and $n_{ah}$) parametrize the scale dependence of $P_B$ ($P_{aB}$) for low and high wavenumbers, respectively.

In the following, we consider a helical magnetic field with $P_{B*} = P_{aB*}$ as a representative case.
It is useful to introduce the characteristic magnetic-field strength at the peak scale $k_*$, defined as~\cite{Durrer:2013pga}
\begin{align}
B_* \equiv \sqrt{\frac{k_*^3 P_{B*}}{\pi^2}}~.
\end{align}
We also define the corresponding characteristic length scale as $\lambda_* \equiv 2\pi/k_*$.
These quantities are used to normalize the left vertical axes in Fig.~\ref{fig:abx}.
In the same figure, the upper horizontal axes are shown in terms of $\Pi d$, while the lower horizontal axes are converted to the photon angular frequency $\omega$ using $\Pi d = m^2 d/(2\omega)$ for conversion into a massive particle of mass $m$.

Several key properties of the integrals $\alpha$, $\chi$, $\beta$ and $\psi$ shown in Fig.~\ref{fig:abx} are summarized as follows:
\begin{itemize}
\item 
$k_* d \lesssim 1:$ For small values of $k_*d$, the amplitudes of the integrals $\alpha$, $\chi$, $\beta$, and $\psi$ monotonically decrease as $k_* d$ decreases.
Their behavior changes qualitatively around $\Pi d = \mathcal{O}(1)$.
For $\Pi d \gg 1$, the integrals exhibit the asymptotic scaling $\alpha, \beta, \psi \propto (\Pi d)^{- 5/3}$ and $\chi \propto (\Pi d)^{- 1}$.
For $\Pi d \ll 1$, $\alpha$ and $\psi$ each converge to a constant value, while $\chi$ and $\beta$ are suppressed and scale as $\chi,\beta \propto \Pi d$.

\item
$k_* d \gg 1:$ For large values of $k_*d$, a characteristic scale emerges around $\Pi d \approx k_* d$ for all four integrals.
In the large-$\Pi d $ region ($\Pi d \gg k_* d$), the asymptotic scaling is the same as in the $k_*d\lesssim1$ case, $\alpha, \beta, \psi \propto (\Pi d)^{- 5/3}$ and $\chi \propto (\Pi d)^{-1}$.
In the intermediate region ($1 \ll \Pi d \ll k_* d$), $\alpha$ and $\psi$ dominate over $\chi$ and $\beta$.
Furthermore, $\psi$ changes sign around $\Pi d \sim k_* d$.
Finally, in the small-$\Pi d$ region ($\Pi d \ll 1$), $\alpha$ and $\psi$ each approach a constant value, while $\chi$ and $\beta$ remain suppressed.
\end{itemize}

In this figure, we observe characteristic behavior arising from the fact that these integrals form KK pairs.
In general, a transition from constant to power-law decay induces a peak structure in the Hilbert transform.
A representative example is the Hilbert transform of $h (\Omega) = 1/(1 + \Omega^2)$, which gives ${\mathcal H} [h] (\Pi) = \Pi / (1 + \Pi^2)$ according to Eq.~\eqref{eq:Hilbert}.
Thus, the localized peak and dip structures in $\chi$ and $\beta$ are direct consequences of the KK relations.
The relative signs of $\chi$ (the partner of $\alpha$) and $\beta$ (that of $\psi$) come from the minus sign in Eq.~\eqref{eq:Hilbert_pair}.

\section{Discussion and conclusions}
\label{sec:discussions}

In this work, we investigate the evolution of photon Stokes parameters induced by photon conversion into other light particles.
We show that, when projecting out the additional fields as unobserved degrees of freedom, the photon sector is effectively described by a non-unitary evolution.
In the perturbative regime, the transformation parameters governing the linear evolution of the Stokes parameters are determined by the Hermitian and anti-Hermitian parts of the effective photon Hamiltonian, which are not independent but are related through Kramers--Kronig (KK) relations as a consequence of the path-ordered structure of the evolution.
We derive these transformation parameters explicitly for photon conversion into dilatons, axions, and gravitons in a background magnetic field.
For stochastic Gaussian magnetic fields, we calculate their ensemble averages and show that the averaged transformation parameters can be expressed as convolution integrals of the magnetic-field power spectra with corresponding integral kernels.
Remarkably, the same set of kernels appears in all three photon-particle systems, revealing a common structure underlying photon conversion into dilatons, axions, and gravitons.

In Sec.~\ref{sec:Stokes}, we formulate the evolution of the photon Stokes parameters under a general non-Hermitian transformation, with applications to photon open systems in mind.
We show that the photon evolution is characterized by eight real transformation parameters, $j_\mu$ and $k_\mu$ ($\mu=0,1,2,3$), with $j_0$ corresponding to an overall phase and therefore leaving seven parameters that govern the evolution of the Stokes parameters.
In Sec.~\ref{sec:conversion}, we investigate how the general non-unitary evolution introduced in Sec.~\ref{sec:Stokes} is realized when photons convert into additional unobserved degrees of freedom.
In the perturbative regime, we express the transformation parameters $j_\mu$ and $k_\mu$ in terms of the second-order contributions of the evolution operator of the full system within the Born approximation.
The path-ordered structure of these contributions leads to KK relations between the Hermitian and anti-Hermitian parts of the effective photon Hamiltonian, demonstrating that the corresponding transformation parameters $j_\mu$ and $k_\mu$ are not independent but form KK pairs.
In Sec.~\ref{sec:magnetic}, we apply this framework to photon conversion into light spin-0 (dilaton and axion) or spin-2 (graviton) degrees of freedom in the presence of an external magnetic field, and explicitly derive the eight transformation parameters for each case in terms of the magnetic-field configuration.
For photon conversion into spin-0 particles, all the relevant transformation parameters are nonvanishing:
$k_\mu$ satisfy a specific identity in each system given by Eqs.~\eqref{eq:iddilaton} and \eqref{eq:idaxion}, while $j_\mu$ satisfy an associated integrated identity \eqref{eq:iddilaton_int} and \eqref{eq:idaxion_int} through the KK relations.
In contrast, no analogous identities arise for conversion into spin-2 particles.

In Sec.~\ref{sec:random}, we consider Gaussian stochastic magnetic fields with possible helical components and study the evolution of both the full system and the photon Stokes parameters in these backgrounds.
We express the ensemble-averaged evolution of the full system over magnetic field realizations to all orders in the interaction, and then obtain the corresponding transformation parameters for each system.
Even in the absence of magnetic helicity, $\alpha\propto\langle k_0\rangle$ and $\chi\propto\langle j_0\rangle$ remain nonzero: the former describes the overall attenuation of all the Stokes parameters due to photon conversion, while the latter represents an overall phase with no effect on them.
Magnetic helicity gives rise to two additional contributions, $\beta\propto\langle k_2\rangle$ and $\psi\propto\langle j_2\rangle$, which describe a change in circular polarization and birefringence, respectively.
We further show that the averaged transformation parameters can be expressed as convolution integrals of the magnetic-field power spectra with corresponding integral kernels.
The same set of kernels appears in all systems, and thus, the effects described by $\alpha$, $\chi$, $\beta$, and $\psi$ arise irrespective of whether photons convert into dilatons, axions, or gravitons, and they form KK pairs.
Consequently, conversion-induced birefringence in a helical magnetic field is necessarily associated with a circular-polarization response; in particular, circular polarization is induced even if initially absent when birefringence occurs.
We further show that, in the long-propagation limit, the magnetic helicity density can be obtained from $\psi$, or equivalently from the frequency integral of $\beta$ through the KK relation.
This provides a way to probe magnetic helicity through circular polarization even for initially unpolarized photons, for which birefringence does not appear.

In Sec.~\ref{sec:numerical}, we adopt a concrete model for the magnetic-field power spectra and investigate the parameter dependence of the ensemble-averaged transformation parameters.
We first examine the convolution kernels and then numerically evaluate the integrals $\alpha$, $\chi$, $\beta$, and $\psi$ as functions of $\Pi d$ for various values of $k_*d$, with the momentum difference $\Pi = \Pi_\varphi - \Pi_\gamma = m^2 / (2 \omega)$, the light particle mass $m$, the photon angular frequency $\omega$, the propagation distance $d$, and the typical wavenumber of the magnetic field $k_*$.
For $k_*d \lesssim 1$, the integrals change their behavior around $\Pi d = {\mathcal O} (1)$, approaching constant or suppressed values at small $\Pi d$ and power-law behavior at large $\Pi d$.
For $k_*d \gg 1$, a characteristic scale appears at $\Pi d \sim k_*d$: $\alpha$ and $\psi$ exhibit sharp transitions at this scale, while $\chi$ and $\beta$ develop localized peak or dip structures.
We find that these behaviors are related to each other through the KK relations between the pairs $(\alpha, \chi)$ and $(\psi, \beta)$.

We conclude by mentioning several possible directions for future work.
First, it would be interesting to extend our formulation to photon conversion in the presence of a background field of the conversion partner.
An important example is photon conversion in an axion dark-matter background~\cite{Masaki_2020}, where a nontrivial axion background may induce qualitatively different evolution of the photon Stokes parameters.
Second, since we have shown that photon conversion in helical magnetic fields can induce birefringence through the parameter $j_2$, it would be interesting to investigate whether this mechanism can contribute to the birefringence of the cosmic microwave background~\cite{minami2020simultaneousdeterminationcosmicbirefringence}.
Finally, a natural extension is to consider cases in which the degrees of freedom into which photons convert exhibit quantum or thermal fluctuations.
Such a formulation would provide a connection to quantum open systems, in which the photon sector is regarded as the system and the conversion partner as the environment.
This extension may allow us to explore how information on quantum or thermal fluctuations of the environment is encoded in the effective evolution of the photon system.
We leave these studies for future work.

\begin{acknowledgments}

The work of W.C. was supported by JST SPRING, Grant Number JPMJSP2148.
R.J. was supported by JSPS KAKENHI Grant Number 24K07013.
K.N. was supported by JSPS KAKENHI Grant Numbers JP24KJ0117 and JP25K17389.

\end{acknowledgments}

\appendix

\section{Expressions for the convolution kernels}
\label{app:A}
In this appendix we summarize the expressions for the convolution kernels in Eq.~\eqref{eq:C}:
\begin{align}
\mathcal{C}_\alpha (y = kd; x = \Pi d)
&=
- \frac{1}{y (x - y z)}
\left[
2 x (x-y z) (\text{Ci}(| x-y z| ) - \ln (| x-y z| ))+\left(x^2+y^2\right) (x-y z) \text{Si}(x-y z)
\right.
\nonumber \\
&\quad
\left.
+(x^2+y^2) \cos (x-y z)-(x-y z) \sin (x-y z)-x^2-xy z-y^2+y^2 z^2
\right]
\Big|_{z = -1}^{z = 1}
~,
\\
\mathcal{C}_\chi (y = kd; x = \Pi d)&=\frac{1}{y}\left[-(x^2+y^2)\text{Ci}(|x-yz|)+(x^2+y^2)\frac{\sin(x-yz)}{x-yz}+2x\,\text{Si}(x-yz)+(x^2+y^2)\ln|x-yz|
\right.
\nonumber \\
&\quad
\left.
+\frac12(x-yz)^2-2x(x-yz)+\cos(x-yz)\right]\Big|_{z = -1}^{z = 1}~.
\\
\mathcal{C}_\beta (y = kd; x = \Pi d)
&=
-\frac{2}{x-y z}
\left[
(x-y z) (\text{Ci}(| x-y z| )-\ln (| x-y z| ))+x (x-y z) \text{Si}(x-y z)+x \cos (x-y z)-x
\right]
\Big|_{z = -1}^{z = 1}
~,
\\
\mathcal{C}_\psi (y = kd; x = \Pi d)
&=2\left[-x\,\operatorname{Ci}(|x-yz|)+x\frac{\sin(x-yz)}{x-yz}+\operatorname{Si}(x-yz)+x\ln|x-yz| -(x-yz) \right]
\Big|_{z = -1}^{z = 1}
~.
\end{align}

\bibliography{main}

\end{document}